\documentclass[manuscript]{aastex}

\usepackage{amsmath}
\usepackage{amssymb}
\usepackage{graphics}
\usepackage[titletoc,title]{appendix}

\def\Lbol{$\textit{L}_\textrm{bol}$}
\def\Tbol{$\textit{T}_\textrm{bol}$}
\def\Trot{$\textit{T}_\textrm{rot}$}
\def\Eup{$\textit{E}_\textrm{up}$}
\def\Lsun{$\textit{L}_{\odot}$}
\def\Msun{$\textit{M}_{\odot}$}
\def\Rsun{$\textit{R}_{\odot}$}
\def\TotalN{$\mathcal{N}$}

\def\G0{$\textit{G}_\textrm{0}$}

\shorttitle{gss30}
\shortauthors{Je et al.}
\begin{document}

\title{``Dust, Ice, and Gas In Time'' (DIGIT) \textit{Herschel} Observations of GSS30-IRS1 in Ophiuchus}

\author{Hyerin Je$^1$, Jeong-Eun Lee$^{1,2}$, Seokho Lee$^1$, Joel D. Green$^2$, and Neal J. Evans II$^2$} 

\affil{$^1$ School of Space Research, Kyung Hee University, Yongin-Si, Gyeonggi-Do 446-701, Republic of Korea; hyerinje@khu.ac.kr, jeongeun.lee@khu.ac.kr }	   

\affil{$^2$ Department of Astronomy, University of Texas at Austin, 2515 Speedway, Stop C1400, Austin, TX 78712-1205, USA}

\begin{abstract}
 As a part of the ``Dust, Ice, and Gas In Time'' (DIGIT) key program on \textit{Herschel}, we observed GSS30-IRS1, a Class I protostar located in Ophiuchus (\textit{d} = 120 pc), with \textit{Herschel}/Photodetector Array Camera and Spectrometer (PACS). More than 70 lines were detected within a wavelength range from 50 $\mu$m to 200 $\mu$m, including CO, H$_{2}$O, OH, and two atomic [O I] lines at 63 and 145 $\mu$m. The [C II] line, known as a tracer of externally heated gas by the interstellar radiation field, is also detected at 158 $\mu$m. All lines, except [O I] and [C II], are detected only at the central spaxel of 9$\arcsec$.4 $\times$ 9$\arcsec$.4. The [O I] emissions are extended along a NE-SW orientation, and the [C II] line is detected over all spaxels, indicative of external PDR. The total [C II] intensity around GSS30 reveals that the far-ultraviolet radiation field is in the range of 3 to 20 \G0, where \G0 is in units of the Habing Field, 1.6 $\times$ 10$^{-3}$ erg cm$^{-2}$ s$^{-1}$. This enhanced external radiation field heats the envelope of GSS30-IRS1, causing the continuum emission to be extended, unlike the molecular emission. The best-fit continuum model of GSS30-IRS1 with the physical structure including flared disk, envelope, and outflow shows that the internal luminosity is 10 \Lsun, and the region is externally heated by a radiation field enhanced by a factor of 130 compared to the standard local interstellar radiation field.
 
\end{abstract}

\section{INTRODUCTION}
 The shocked gas induced by interaction between the protostellar jet and the surrounding interstellar medium cools through copious molecular emission lines and forbidden atomic lines, such as CO, H$_{2}$O, OH, and [O I] lines, in the far infrared (FIR) range \citep{Nisini02}. \citet{Giannini01} suggested that the FIR lines are good tracers of the energy budget related to outflows based on the observations of young stellar objects (YSOs) with the Long Wavelength Spectrometer (LWS) on board the \textit{Infrared Satellite Observatory} (\textit{ISO}). According to their results, the powerful outflows of Class 0 sources are the results of larger envelope mass infall rates as compared with more evolved sources, indicating that outflow can be an evolutionary parameter of YSOs. Therefore, the observations in the FIR band are important to understand the nature of the early stage YSOs.
 
  The [O I] and [C II] lines at FIR wavelengths have been used to measure the strength of the interstellar radiation field (ISRF) in order to predict the presence of photo-dissociation regions (PDRs) as well as understand the heating mechanism associated with YSOs \citep{Yui93, Nisini02}. Although the enhanced interstellar radiation fields result in the high excitation conditions, the effects of the external radiation fields on the low mass YSOs are poorly understood.

  Observations of YSOs using \textit{ISO}/LWS stimulated studies of origins of the FIR emission as well as line cooling budgets in atomic and molecular line emission \citep{Giannini01, Nisini02}. However, the \textit{ISO} beam was too big to discriminate the different gas conditions. Recently, large numbers of embedded YSOs have been observed \citep{Green13, Karska13} with the \textit{Herschel Space Observatory} \citep{Pilbratt10} with much better spatial and spectral resolutions and much better sensitivity. Many weak molecular and atomic lines have been detected with the Photodetector Array Camera and Spectrometer \citep[PACS; ][]{Poglitsch10} on board \textit{Herschel}. Here we present the detailed analysis of line and continuum fluxes of GSS30-IRS1, which was observed by the \textit{Herschel}/PACS instrument as part of the \textit{Herschel} Key Program, DIGIT, ``Dust, Ice, and Gas in Time'' (PI: N.Evans).
  
   The Ophiuchus (hereafter Oph) molecular cloud, at a distance of 120 $\pm$ 4 pc \citep{Loinard08}, is a good region for studying the impact of irradiation fields because the interstellar radiation field impinging on the Oph cloud is stronger than the average for our solar neighborhood \citep{Liseau99}. \citet{Lindberg14} showed that the external radiation field could affect the physical and chemical properties around YSOs based on \textit{Herschel} observations.
  
\begin{figure*}
 \includegraphics[width=1 \textwidth]{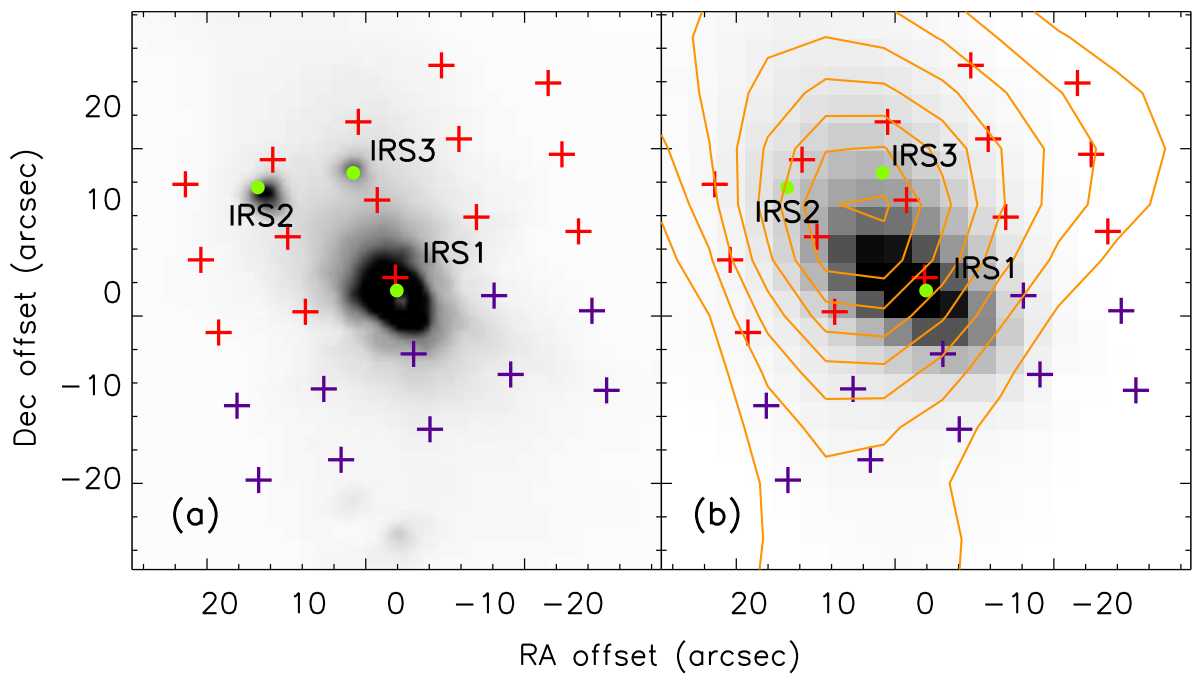}
 \centering
 \caption{\textit{Spitzer}-IRAC 2 (4.5 $\mu$m) image (a) and \textit{Herschel}/PACS (70 $\mu$m) image (b) of GSS30. The purple and red crosses represent the positions of PACS pixels. The spaxels where we used for measuring the [O I] fluxes are represented as red crosses. The green circles show the locations of YSOs. The orange contours for the SCUBA 850 $\mu$m data are overplotted in the right panel. Contour levels are increasing in 10, 20, 30, 40, 50, 60, 70, 80, and 90 \% of the peak flux.}
 \label{Spitzer_PACS} 
\end{figure*}

   One of the well studied targets, GSS30, is a bipolar reflection nebula in the Oph molecular cloud. \citet{Weintraub93} reported that there are three infrared sources: GSS30-IRS1, -IRS2, and -IRS3 (Figure~\ref{Spitzer_PACS}a)\footnote{The IRAC image was downloaded from the \textit{Spitzer} data archive.}. According to the geometry of the infrared emission, GSS30-IRS1 is the illuminating source in the nebula and the strongest source in the near-infrared \citep{Castelaz85, Weintraub93, Zhang97} (Figure~\ref{Spitzer_PACS}b). GSS30-IRS1 has been classified as a Class I YSO by its spectral energy distribution (SED) \citep{Grasdalen73, Elias78, Wilking89} with \Lbol{} of 14.5 \Lsun{} and \Tbol{} of 138 K \citep{Green13}. IRS2 has no detected emission at millimeter wavelength, indicating that it is a more evolved (T-Tauri) star \citep{Andre94}. IRS3 was detected at 2.7 mm \citep{Zhang97} and 6 cm \citep{Leous91}. It has been known as a Class I YSO with \Lbol{} of 0.13 \Lsun{} \citep{Bontemps01}, much fainter than IRS1. 
  
  The origin of the associated molecular outflow is unclear. \citet{Tamura90} detected high velocity millimeter CO 1--0 line emission in this region of roughly 5$\arcmin$ in extent. However, both the blue- and red-shifted lobes lie to the south of GSS30-IRS1, deviating from the bipolar morphology seen in the reflection nebula. They interpreted this asymmetry of the CO outflow lobes in terms of the interaction between one side of the lobe and a dense NH$_{3}$ cloud core which is located to the northeast of GSS30-IRS1. \citet{Andre90} suggested that high velocity gas components reported by \citet{Tamura90} are associated with the VLA1623 jet, which is located southeast of GSS30. Recently, \citet{Jorgensen09} detected wing emission clearly associated with the GSS30-IRS1 in the HCO$^{+}$ lines observed with the Submillimeter Array (SMA).   
 
 In order to study the effect of external radiation fields on GSS30-IRS1, we analyze the observed PACS atomic line spectra and model the continuum emission using the continuum radiative transfer code (RADMC-3D). For molecular lines associated directly with the protostar, the rotational diagrams of molecular transitions and a non-LTE large velocity gradient (LVG) code (RADEX) have been utilized. The outline of this paper is as follows: in \S~\ref{sec:obs}, we describe the \textit{Herschel} observations and data reduction. The observational results are presented in \S~\ref{sec:result}. The detailed analyses, including line and the continuum emission, are described in \S~\ref{sec:line} and \S~\ref{sec:continuum}. We also discuss the effects of external heating in \S~\ref{sec:continuum}, and summarize our results in \S~\ref{sec:summary}, respectively.

\section{OBSERVATIONS AND DATA REDUCTION} \label{sec:obs}
 GSS30-IRS1 was observed on 10 March 2011 ($\lambda$ $\sim$ 50--75 and 100--145 $\mu$m; AOR: 1342215678, $\lambda$ $\sim$70--100 and 140--210 $\mu$m; AOR: 1342215679) by \textit{Herschel}/PACS in a pointed range-scan spectroscopy mode. The PACS instrument consists of a 5 $\times$ 5 array whose unit spatial pixel (hereafter spaxel) covers 9$\arcsec$.4 $\times$ 9$\arcsec$.4 with a spatial resolution of $\sim$ 9$\arcsec$ at the shorter wavelengths and $\sim$ 18$\arcsec$ at longer wavelengths. Its spectral coverage ranges from 50 $\mu$m to 210 $\mu$m with a spectral resolution of $\lambda$/$\delta\lambda\sim$ 1000--3000, and its coverage is divided into four segments, covering $\lambda$ $\sim$ 50--75, 70--105, 100--145, and 140--210 $\mu$m. The edges of the segments ($\lambda$ $<$ 55 $\mu$m and 95 $\leq$ $\lambda$ $\leq$ 103 $\mu$m) suffer from noise, and the spectral fluxes appear much higher than their real fluxes longwards of 190 $\mu$m. For those reasons, we limited the wavelength ranges to 55 $\leq$ $\lambda$ $<$ 95 $\mu$m and 103 $\leq$ $\lambda$ $<$ 190 $\mu$m in this study to calculate the line fluxes accurately. The sky background and telescope emission were subtracted using two nod positions located 6$\arcmin$ from the source in opposite directions. 

 The basic data reduction of the DIGIT observations was performed with \textit{Herschel} Interactive Processing Environment (HIPE) v.13 \citep{Ott10}. The complete description for the data reduction can be found in Green et al. (in prep.). This reduction provided the best signal-to-noise ratio (S$/$N) and better absolute flux calibration. In order to measure the total line fluxes, we used the two different methods due to the spatial distributions of molecular and atomic lines. For all molecular lines, the spectra extracted from the central spaxel were corrected for the point spread function (PSF) of PACS. However, we did not apply this method to atomic lines because the [O I] and [C II] lines are detected over all 25 spaxels. In addition, the PACS fluxes for the atomic lines are contaminated by the emission in the sky positions (Sect.~\ref{sec:nod_position}). Thus, we measured the fluxes from the nod B observation (with off position 6$\arcmin$ to the south from GSS30-IRS1),  in which [O I] and [C II] appear in emission. While the [C II] lines are analyzed in all 25 PACS spaxels, the [O I] fluxes are measured over only 15 spaxels (red crosses in Figure~\ref{Spitzer_PACS}) selected to avoid the effects of the outflow by VLA1623 in the southwest spaxels (see Sect.~\ref{sec:spectrum_line} and Sect.~\ref{sec:nod_position}).
 
  All lines were fitted by the Spectroscopic Modeling Analysis and Reduction Tool \citep[SMART; ][]{higdon04}, and the Leiden Atomic and Molecular Database (LAMDA) was used as the molecular database \citep{Schoier05}.

\section{RESULTS} \label{sec:result}
 \subsection{Spatial Distribution of Continuum and Line Emission} \label{sec:distribution}
\begin{figure*}
 \includegraphics[scale=0.6]{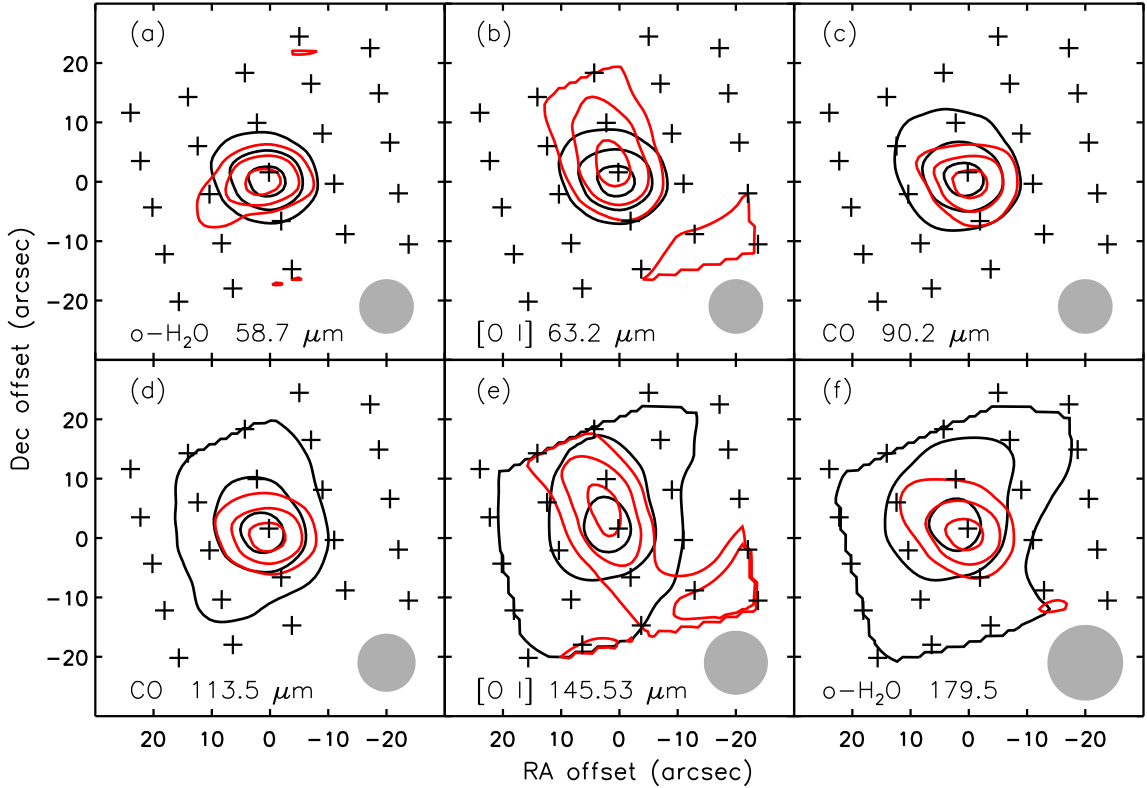}
 \centering
 \caption{Contour maps of selected line flux (red solid line) and local continuum (black solid line). The positions of PACS pixels are represented as the black crosses. In each map, contour levels are 30, 60, and 90 \% of the peak flux of each line. The gray filled circles indicate the beam at each wavelength. The wavelength of each line transition is presented in the bottom of each box. The local continuum bands are 59--61, 63--65, 88--90, 177--179, 145--147, and 115--117 $\mu$m from the top left in the clockwise direction. The extended line structure seen in the SW of GSS30-IRS1 is caused by the outflow of VLA1623 (Sect.~\ref{sec:spectrum_line}). Unlike other molecular lines, the [O I] line is extended toward northeast direction. All local continuum are more extended than lines, except for [O I]. }
 \label{contour_line_conti} 
\end{figure*}

 One interesting observational result is that the continuum emission extends to the northeast. The same feature has been shown as scattered light in an \textit{H}-band image from the \textit{Very Large Telescope} (\textit{VLT}) \citep{Chen07} and a 1.6 $\mu$m Hubble Space Telescope (HST) image from the Near Infrared Camera and Multi-Object Spectrometer (NICMOS) \citep{Allen02}. The SCUBA 850 $\mu$m map \citep{Di Francesco08} also shows the same extended structure (Figure~\ref{Spitzer_PACS}b). In Figure~\ref{contour_line_conti}, the continuum emission is significantly more extended than the line emission at longer wavelengths except for [O I], although a few lines at longer wavelengths, $\lambda$ $>$ 120 $\mu$m, appear extended, which is mostly because of the large PACS PSF. Thus, we explore the cause of the extended dust continuum emission by modeling the SED of GSS30-IRS1 in Sect.~\ref{sec:continuum}.

 \subsection{Continuum Deconvolution}
\begin{figure*}
 \includegraphics[width=1 \textwidth]{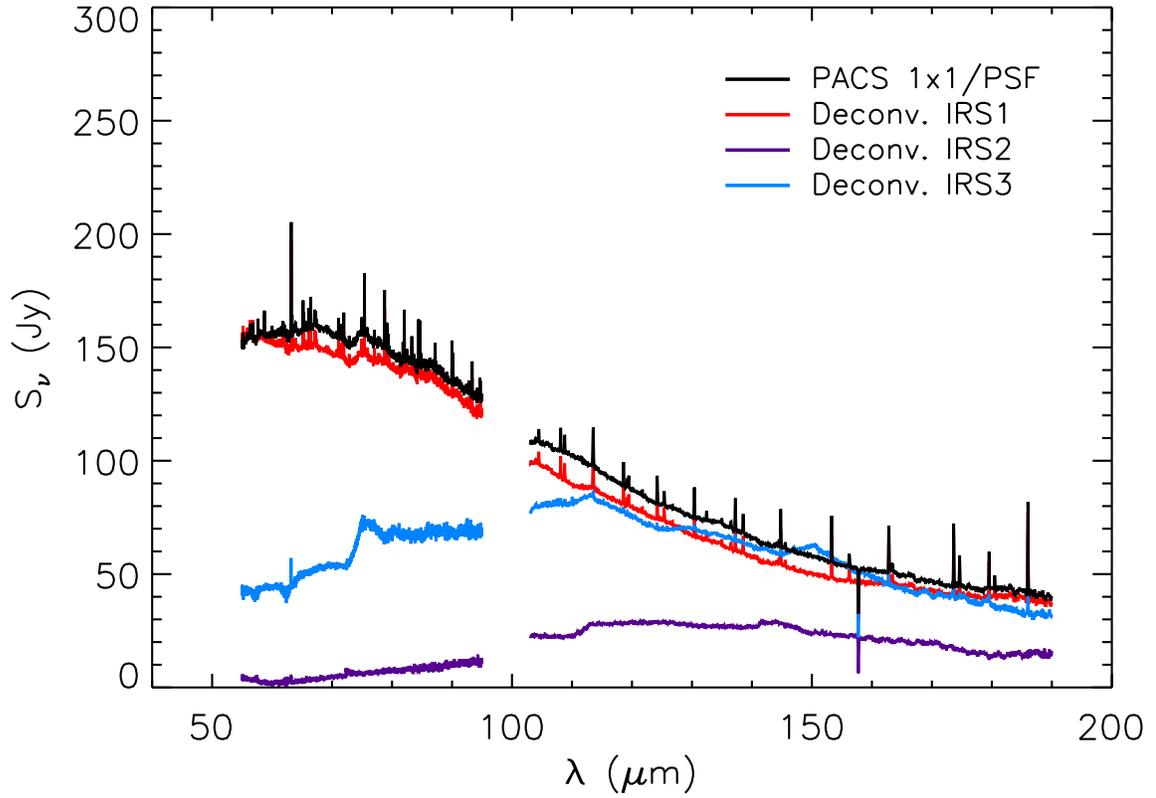} 
 \centering
 \caption{The spectra of the deconvolved IRS1 (red solid line), IRS2 (purple solid line), and IRS3 (blue solid line) YSOs in GSS30. Fluxes are measured from the central spaxel and corrected for PACS/PSF, represented as black solid lines.}
 \label{SED_deconv} 
\end{figure*}

 As seen in Figure~\ref{Spitzer_PACS}a, there are three YSOs within the PACS field of view (FOV). Thus, it is difficult to measure the total flux of IRS1 from the PACS total fluxes because of the contamination by those multiple sources. In addition, the PACS image at 70 $\mu$m (Figure~\ref{Spitzer_PACS}b) shows that the continuum is clearly extended. In order to test whether this extended feature is due to the contribution from three YSOs or the external heating, we decompose the fluxes of IRS1, IRS2, and IRS3 with the assumption that all sources are point-like. The key concept of deconvolution method is the same as used in \citet{Lindberg14}, except that we adopted the PSF from the PACS spectrometer beams version 3\footnote{See https://nhscsci.ipac.caltech.edu/sc/index.php/Pacs/PACSSpectrometerBeams}. Figure~\ref{SED_deconv} shows the deconvolved spectra of three sources in the PACS range. All spectra are deconvolved within the PACS FOV. Thus, we note that deconvolved flux densities of IRS2 and IRS3 are underestimated because they are located near the edge of the PACS FOV. However, the deconvolved flux densities of IRS1 are similar to the fluxes measured from the central spaxel and then corrected for PACS/PSF, within the calibration error (10 $\%$). It indicates that IRS1 is located at the central position of PACS and the deconvolved flux densities are the total flux densities of IRS1. The continuum emission from IRS3 is comparable with that from IRS1 beyond 100 $\mu$m (Figure~\ref{SED_deconv}).
 
\begin{figure*}
 \includegraphics[width=1 \textwidth]{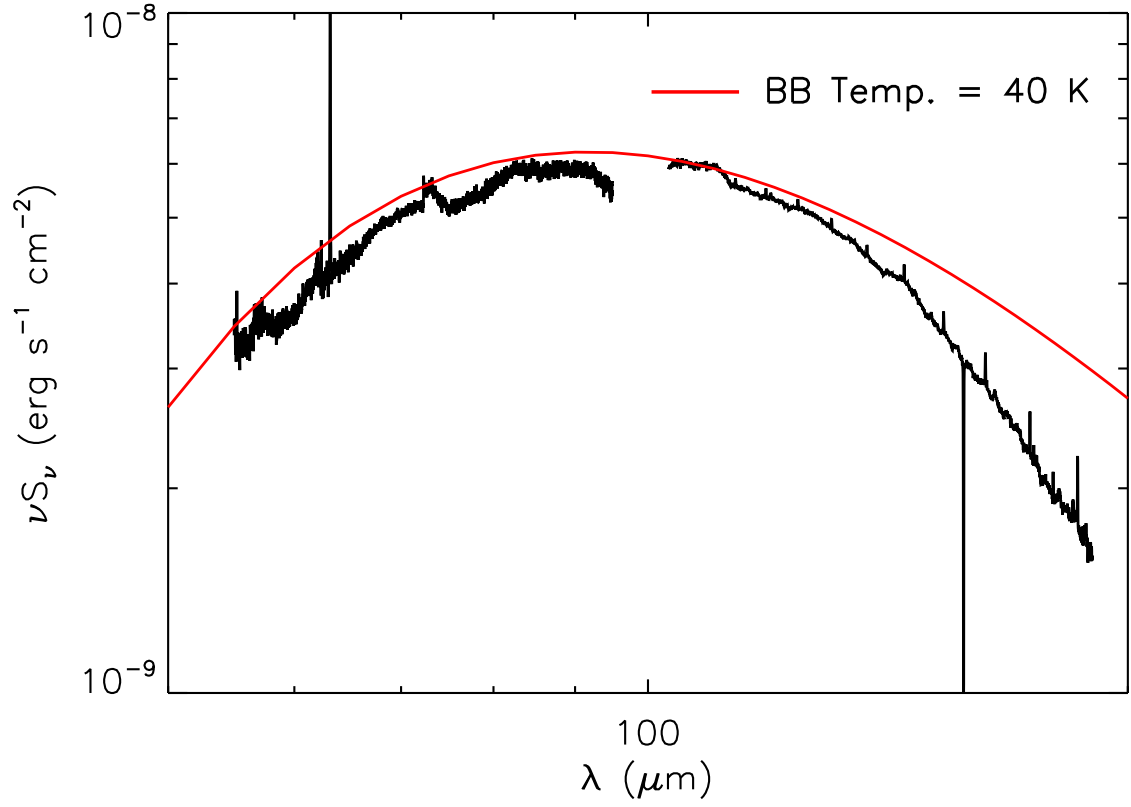}
 \centering
 \caption{The SED of the continuum residual in PACS wavelength range (black solid line) and fitting of black-body dust temperature of 40 K (red solid line).}
 \label{SED_residual} 
\end{figure*}

 Figure~\ref{SED_residual} shows the residual emission after subtracting emission by individual infrared sources from the total fluxes over 5 $\times$ 5 spaxels as described in the equation below:
   \begin{eqnarray} \label{residual}
     F_{\nu, \textrm{ residual}} = F_{\nu, \textrm{ 5$\times$5}} - ( F_{\nu, \textrm{ IRS1, deconv.}} + F_{\nu, \textrm{ IRS2, deconv.}} + F_{\nu, \textrm{ IRS3, deconv.}} ),
   \end{eqnarray}
   
$F_{\nu, \textrm{ 5$\times$5}}$ is the flux measured over the PACS 25 spaxels and $F_{\nu, \textrm{ name, deconv.}}$ are the deconvolved fluxes of each source in the PACS FOV. The residual spectrum is fitted by the black-body temperature of $\simeq$ 40 K.

 \subsection{The Spectrum of GSS30-IRS1} \label{sec:spectrum_line}
\begin{figure*}
 \includegraphics[width=1 \textwidth]{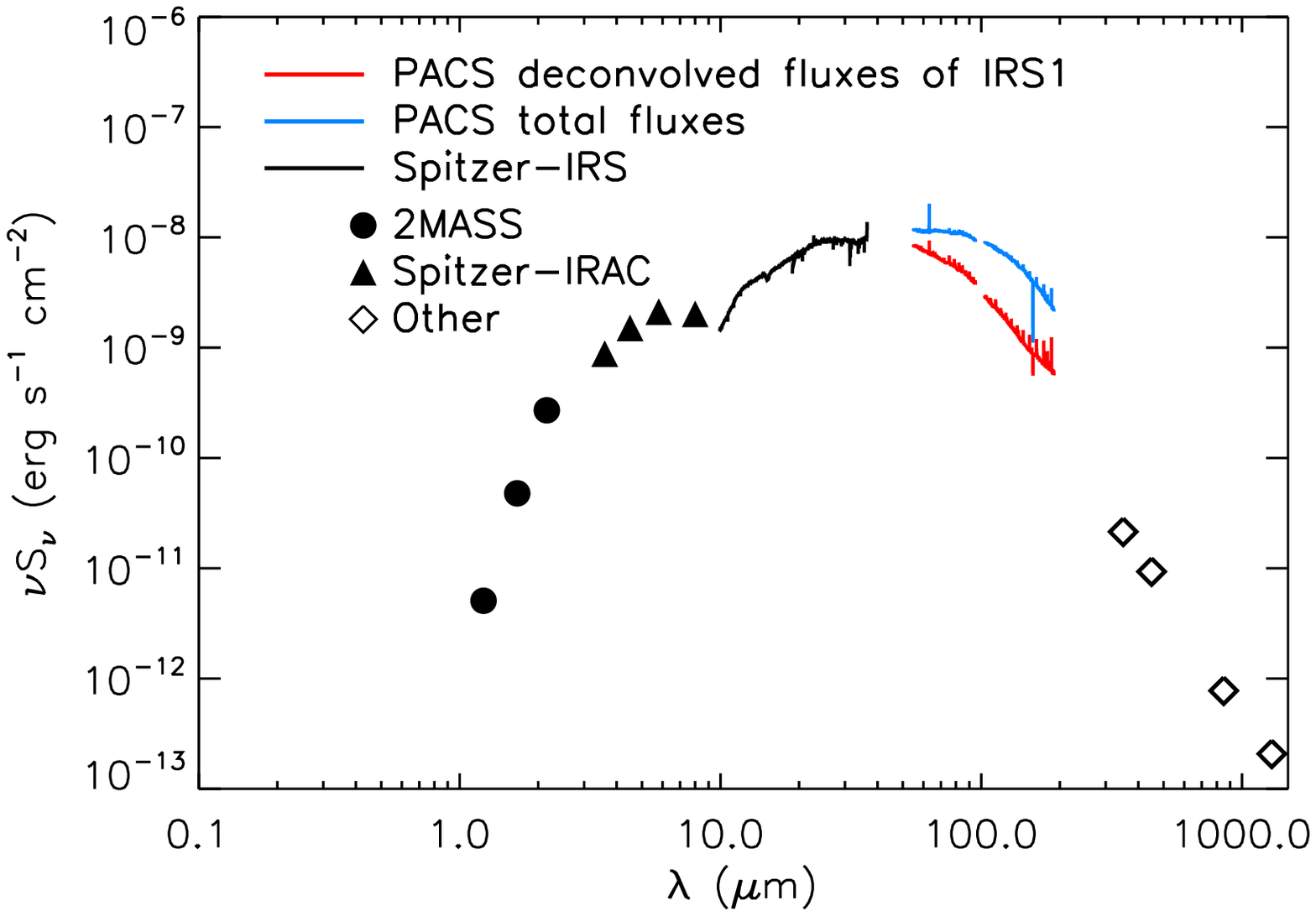}
 \centering
 \caption{SED with \textit{Herschel}/PACS, 2MASS (JHK; filled circles), \textit{Spitzer}-IRAC (3.6, 4.5, 5.8, 80 $\mu$m; filled triangles), and \textit{Spitzer}-IRS (10-36 $\mu$m; black solid line) from the ``c2d'' legacy program \citep{Lahuis06} and the ``IRS$\_$Disk'' GTO program \citep{Furlan06}. The red and blue PACS spectra have been extracted differently; the red spectrum represents the deconvolved IRS1 SED while the blue spectrum shows the SED over the whole 25 PACS spaxels after subtracting the deconvolved fluxes from IRS2 and IRS3. The Sub-mm data are from SHARC-II, SCUBA \citep[350 $\mu$m in 25$\arcsec$ radius, 450 $\mu$m in 9$\arcsec$ beam, 850 $\mu$m in 15$\arcsec$ beam; ][]{van Kempen09}, and IRAM \citep[1.3 mm in 11$\arcsec$ beam; ][]{Motte98}.}
 \label{SED_whole} 
\end{figure*}

 In Figure~\ref{SED_whole}, we present the SED of GSS30-IRS1 up to 1300 $\mu$m obtained from the photometry of 2MASS and \textit{Spitzer}-IRAC and the spectroscopy of \textit{Herschel}/PACS and \textit{Spitzer}-IRS \citep{Lahuis06} from the \textit{Spitzer} Legacy Program ``From Molecular Cores to Planet Forming Disks'' (c2d) \citep{Evans03} and the ``IRS$\_$Disk" Guaranteed Time Observations (GTO) program \citep{Furlan06}. The sub-mm data are from the Submillimeter High Angular Resolution Camera II (SHARC-II), Submillimetre Common-User Bolometer Array (SCUBA) \citep{van Kempen09}, and the Institut de Radioastronomie Millimetrique (IRAM) \citep{Motte98}. The deconvolved SED of IRS1 (Figure~\ref{SED_whole}, red line) exhibits a bolometric luminosity of 8.5 \Lsun{} and a bolometric temperature of 192.8 K. 

\begin{figure*}
 \includegraphics[scale=0.8]{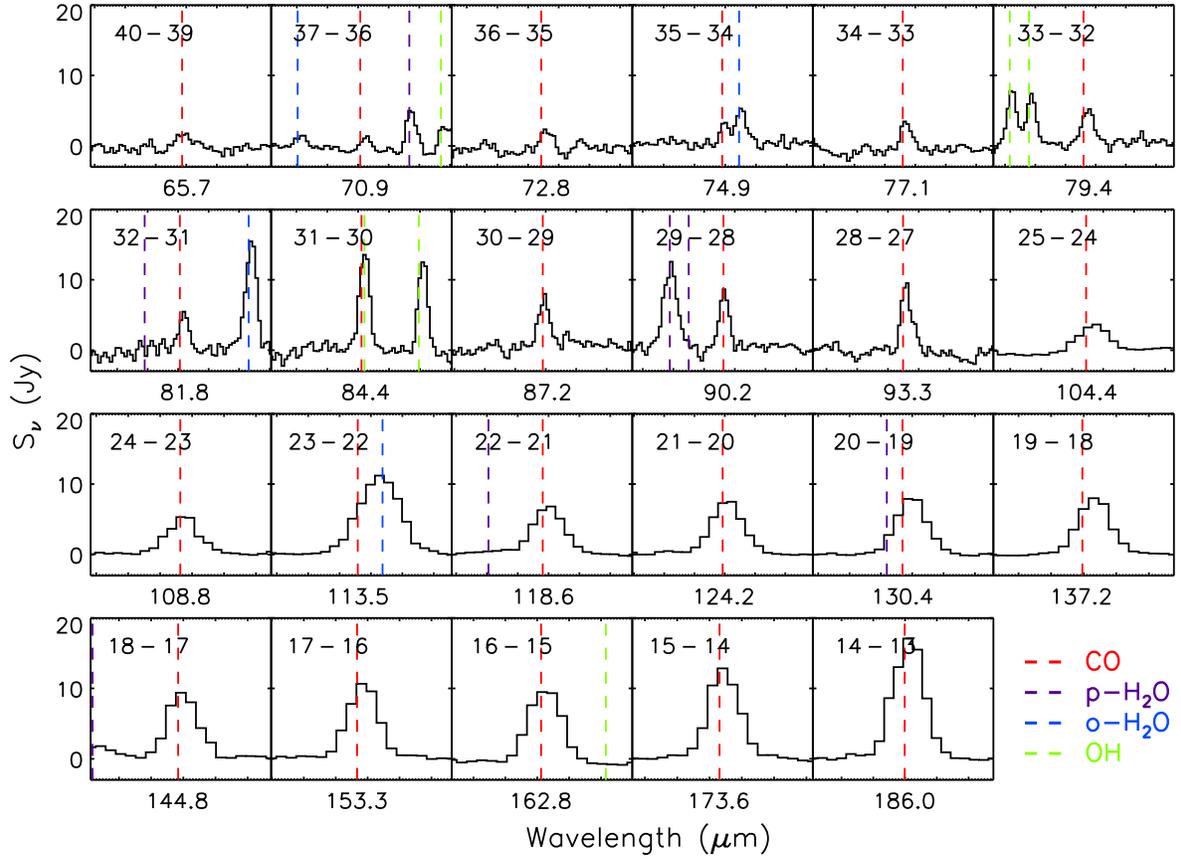}
 \centering
 \caption{Detected CO lines from GSS30-IRS1. Lines are extracted from the central spaxel. The wavelength of each transition is marked: CO (red), p-H$_{2}$O (purple), o-H$_{2}$O (blue), and OH (green). The rotational transition level of each line is listed on top of each box.}
 \label{linemap_CO} 
\end{figure*}

\begin{figure*}
 \includegraphics[scale=0.8]{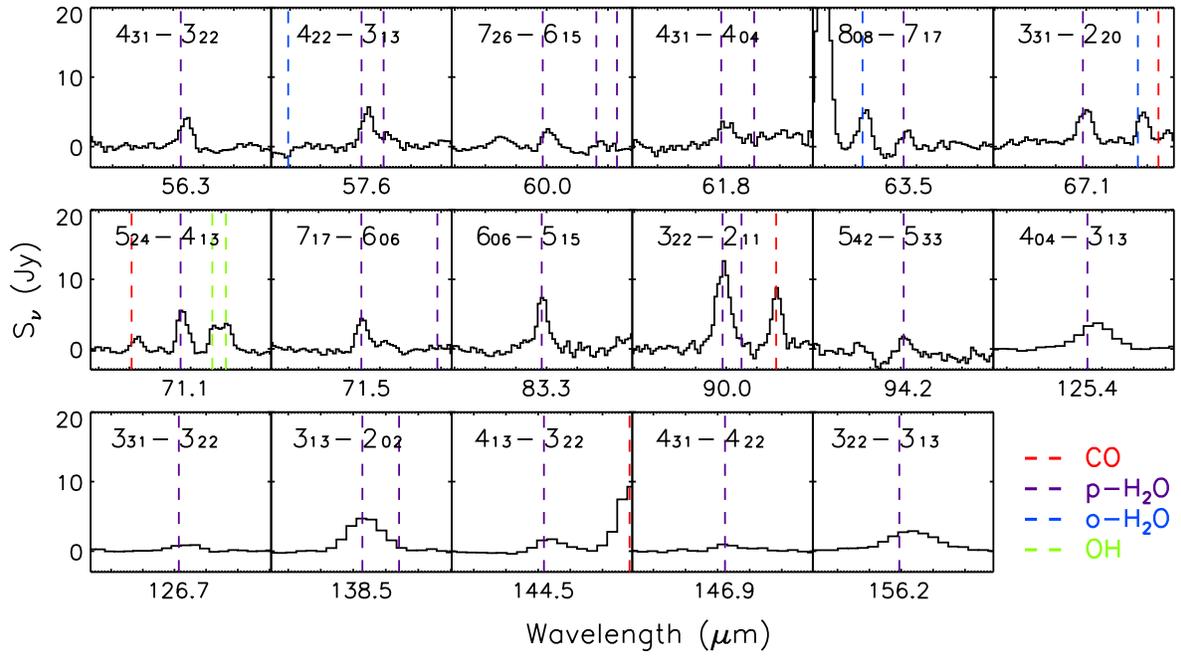}
 \centering
 \caption{The same as Figure~\ref{linemap_CO} but for p-H$_{2}$O lines.}
 \label{linemap_pH2O} 
\end{figure*}

\begin{figure*}
 \includegraphics[scale=0.8]{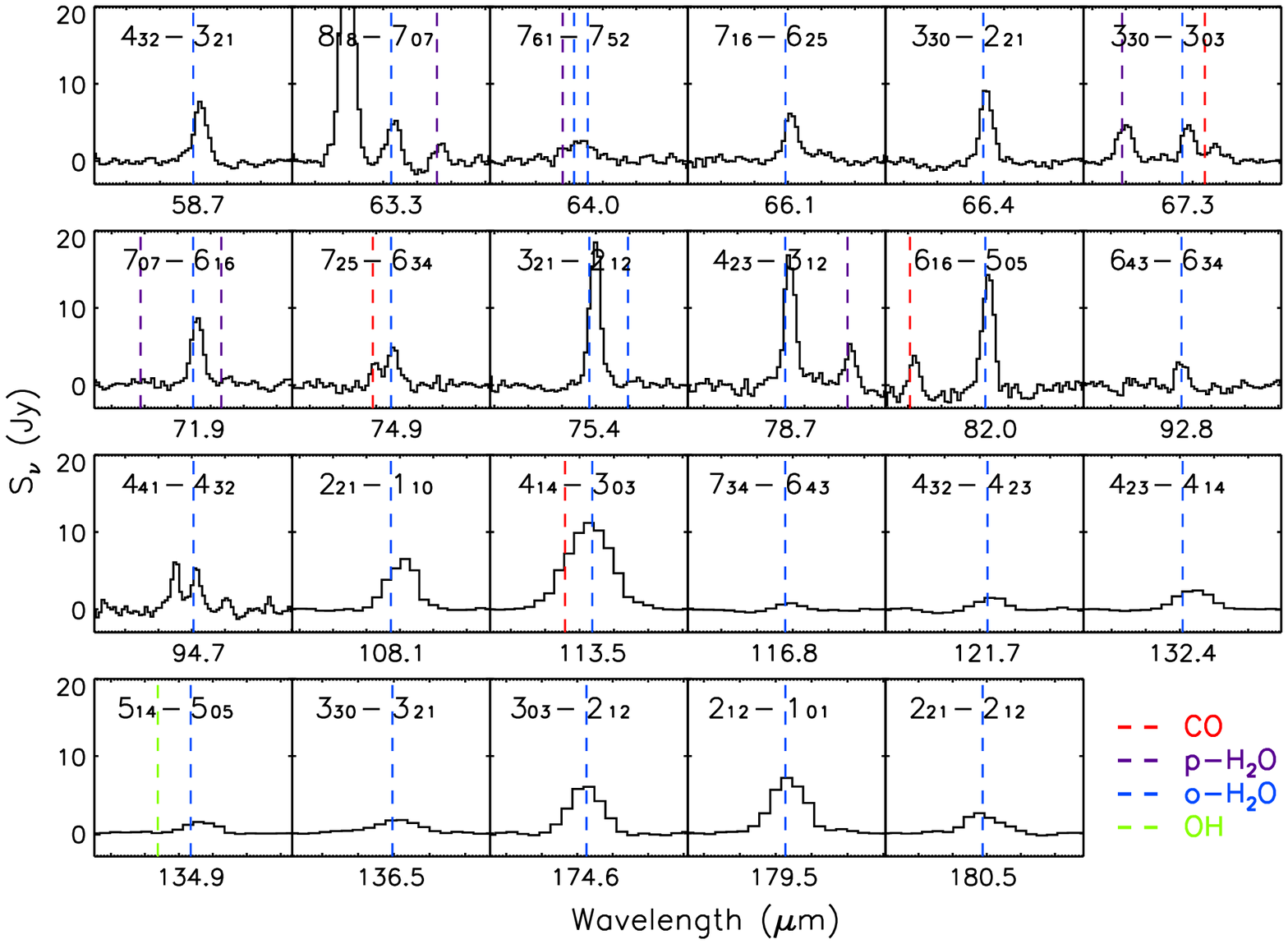}
 \centering
 \caption{The same as Figure~\ref{linemap_CO} and~\ref{linemap_pH2O} but for o-H$_{2}$O lines.}
 \label{linemap_oH2O} 
\end{figure*}

\begin{figure*}
 \includegraphics[scale=0.8]{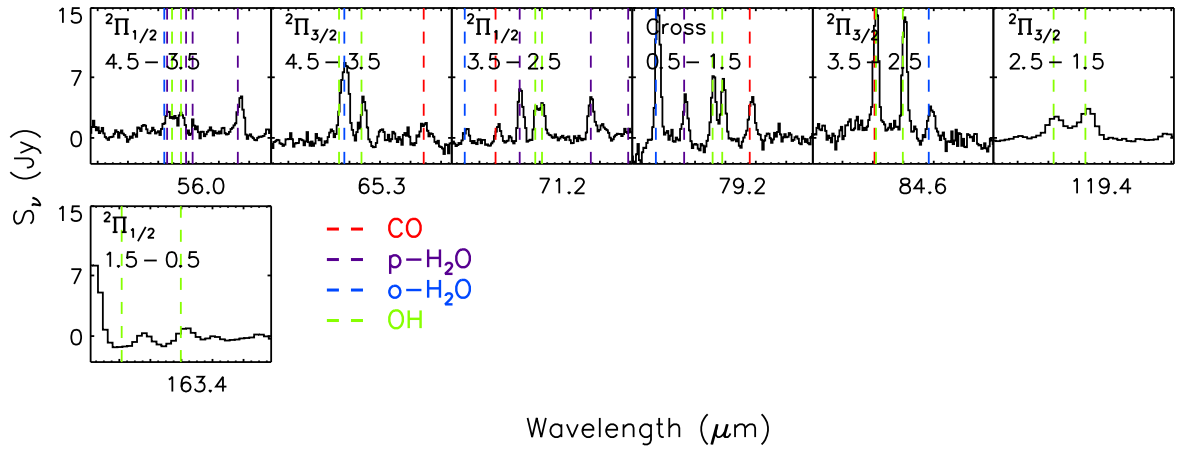}
 \centering
 \caption{The same as Figure~\ref{linemap_CO},~\ref{linemap_pH2O}, and~\ref{linemap_oH2O} but for OH lines.}
 \label{linemap_OH} 
\end{figure*}

 The rich PACS spectra show molecular lines including CO, H$_{2}$O, and OH, and two atomic [O I] lines in emission. However, as seen in Figure~\ref{SED_whole}, the [C II] 158 $\mu$m line has a deep absorption feature. All detected lines are listed in Table~\ref{tbl_flux} with blended line emissions noted. We excluded blended lines in our analyses. Detected lines are presented in Figure~\ref{linemap_CO} to Figure~\ref{linemap_OH}. Although weak CO lines at long wavelengths are detected toward spaxels south of the central source, those lines probably trace the outflow related to a nearby Class 0 source, VLA1623, located southeast (5$\arcsec$ offset in right ascension and 1$\arcmin$ offset in declination from GSS30-IRS1) of GSS30-IRS1 \citep{Dent95, Yu97, Pontoppidan02}.  
Blue and red outflow wings, related to VLA1623 and other HH objects, are clearly seen south of GSS30-IRS1 \citep{Bjerkeli12}. In addition, the $^{12}$CO \textit{J} = 3--2 observations show the outflow emission dominated by VLA1623 toward the south of GSS30-IRS1 \citep{Marel13}. To avoid contamination, we present only the CO, H$_{2}$O, and OH lines extracted from the central spaxel (Figure~\ref{linemap_CO} to Figure~\ref{linemap_OH}). 

 In contrast, two detected atomic lines ([O I] and [C II]) are distributed over a region larger than the central spaxel and they are often affected by strong emission in the nod positions used for sky subtraction. We investigate the spatial distributions of these lines using two different sky subtractions in the next section.

 \subsection{Spatial Distribution of Atomic Lines}  \label{sec:nod_position}
\begin{figure*}
 \includegraphics[scale=0.8]{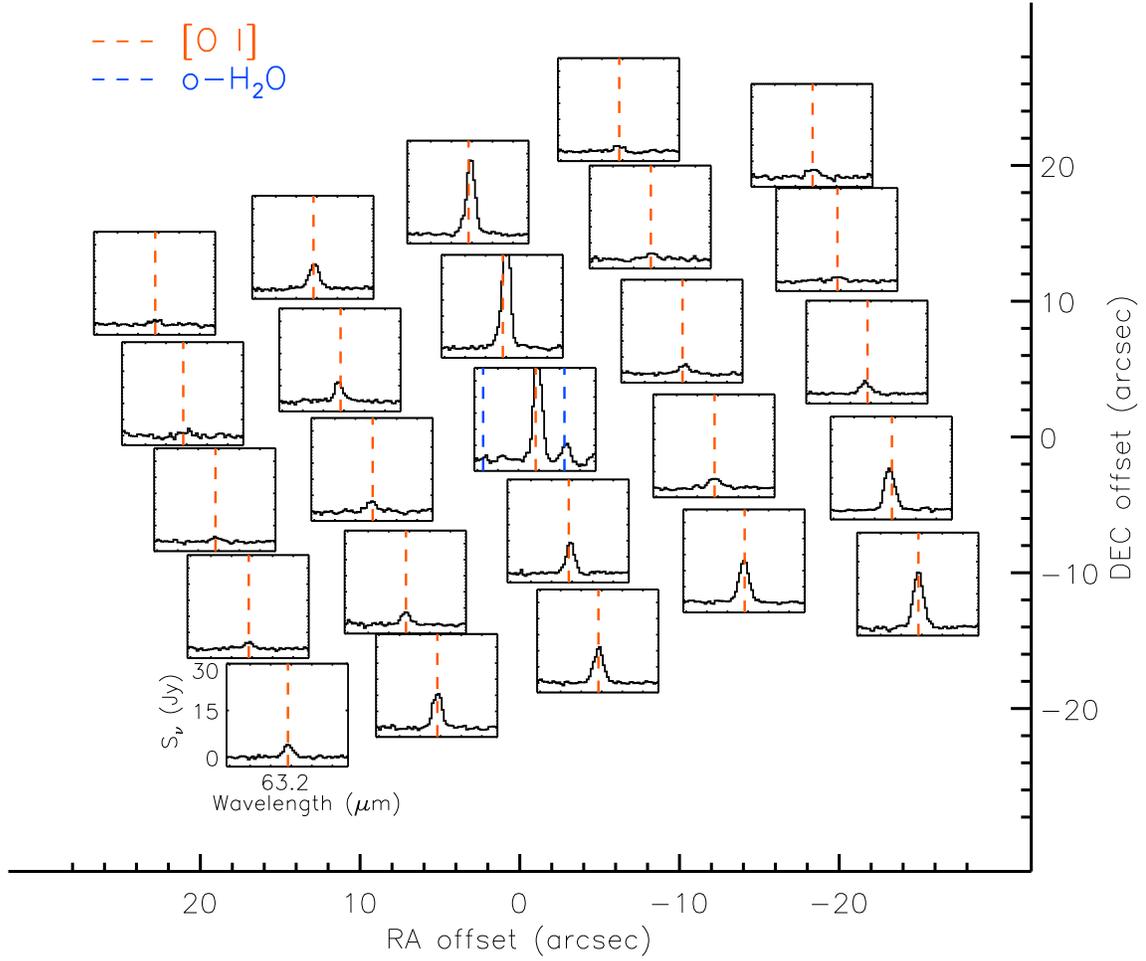}
 \centering
 \caption{The whole PACS spectrum map of [O I] 63 $\mu$m line. The [O I] lines southwest
from the central spaxel are affected by the outflow of VLA1623.}
 \label{linemap_OI63} 
\end{figure*}

 Figure~\ref{linemap_OI63} shows the entire PACS spectral line map of the [O I] 63 $\mu$m line, which is the average of the two nods subtraction. The [O I] line emission is elongated along the NE-SW direction, and the morphology is similar to a part of the outflow cone detected by the SMA HCO$^{+}$ 3--2 and HCN 3--2 observations \citep{Jorgensen09}. This indicates that the [O I] traces the outflow structure (see next section). Although [O I] emission extends into the region where IRS3 is located, we conclude that the [O I] emission is related to IRS1, due to the strong peak at the IRS1 position in our line map. In addition, the outflow tracer HCO$^{+}$ and HCN lines are clearly detected toward IRS1 \citep{Jorgensen09}. There is a separate structure to the SW of GSS30-IRS1 in the line map of the [O I] (Figure~\ref{linemap_OI63}). As mentioned above, the structure seems related to the outflow by VLA1623. 
  
\begin{figure*}
 \includegraphics[width=0.5 \textwidth]{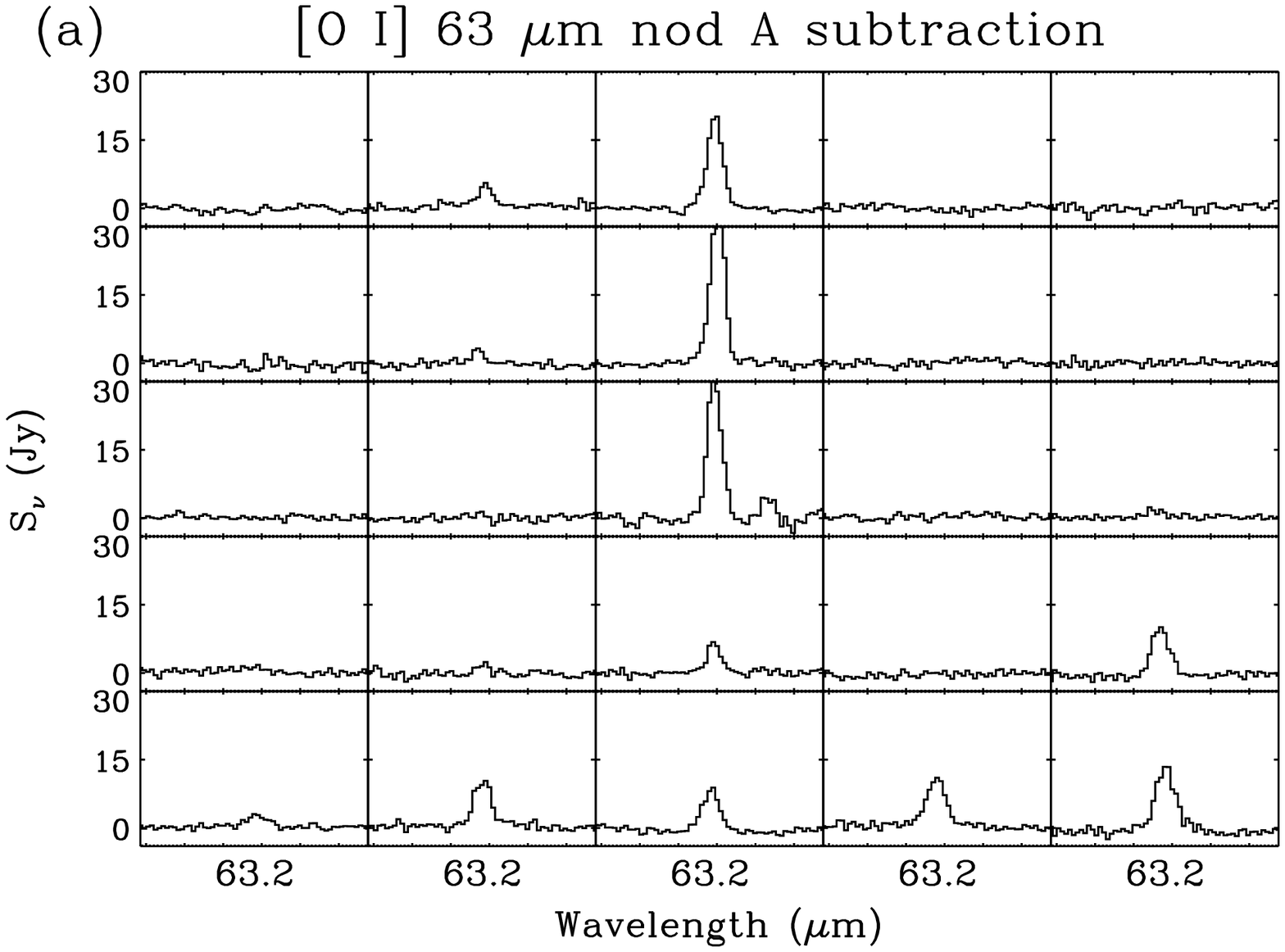}
 \includegraphics[width=0.5 \textwidth]{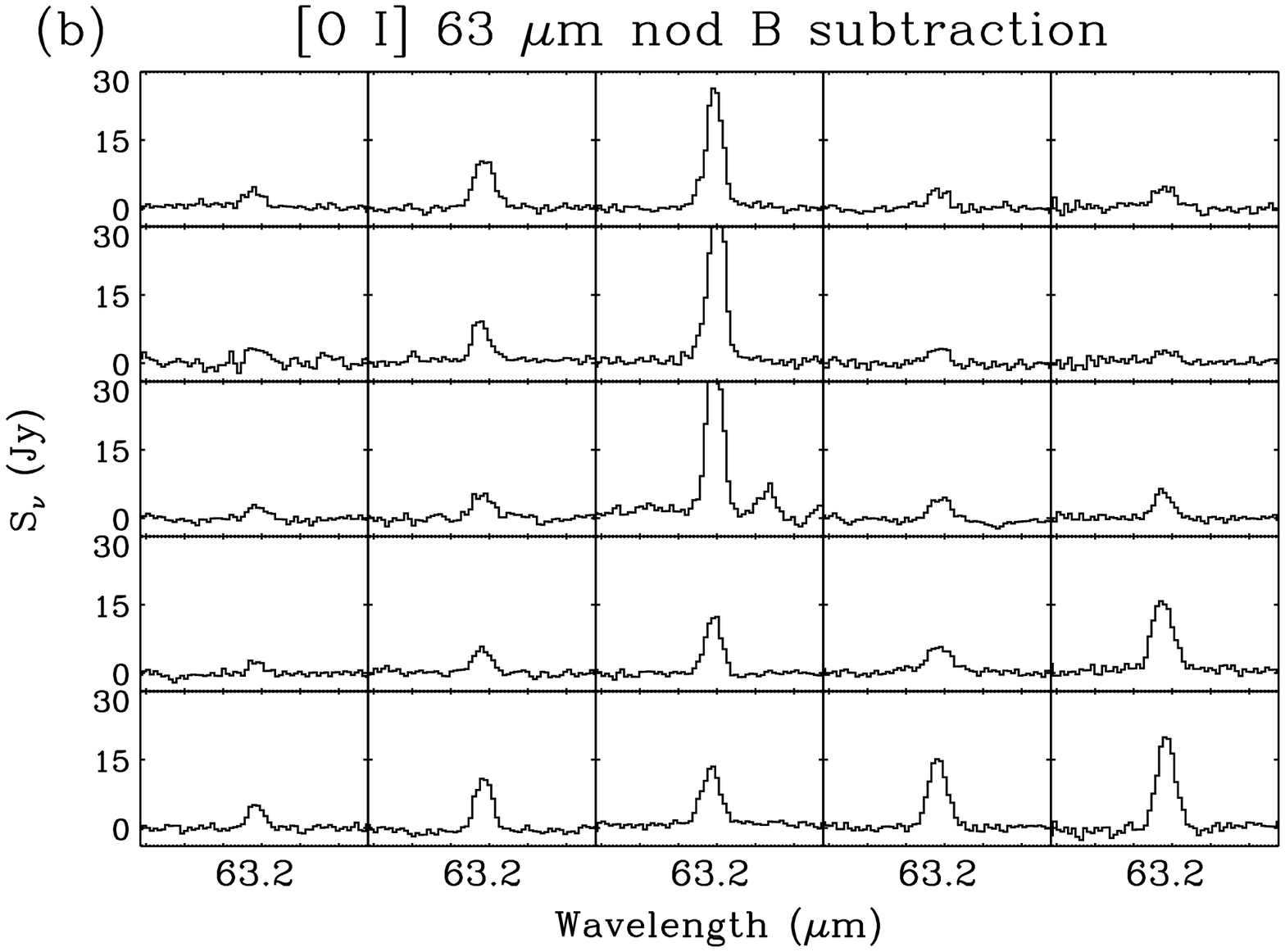} 
 \caption{PACS spectral maps of [O I] 63 $\mu$m, where each nod position spectrum is subtracted.}
 \label{linemap_nod_OI63} 
\end{figure*}

\begin{figure*}
 \includegraphics[width=0.5 \textwidth]{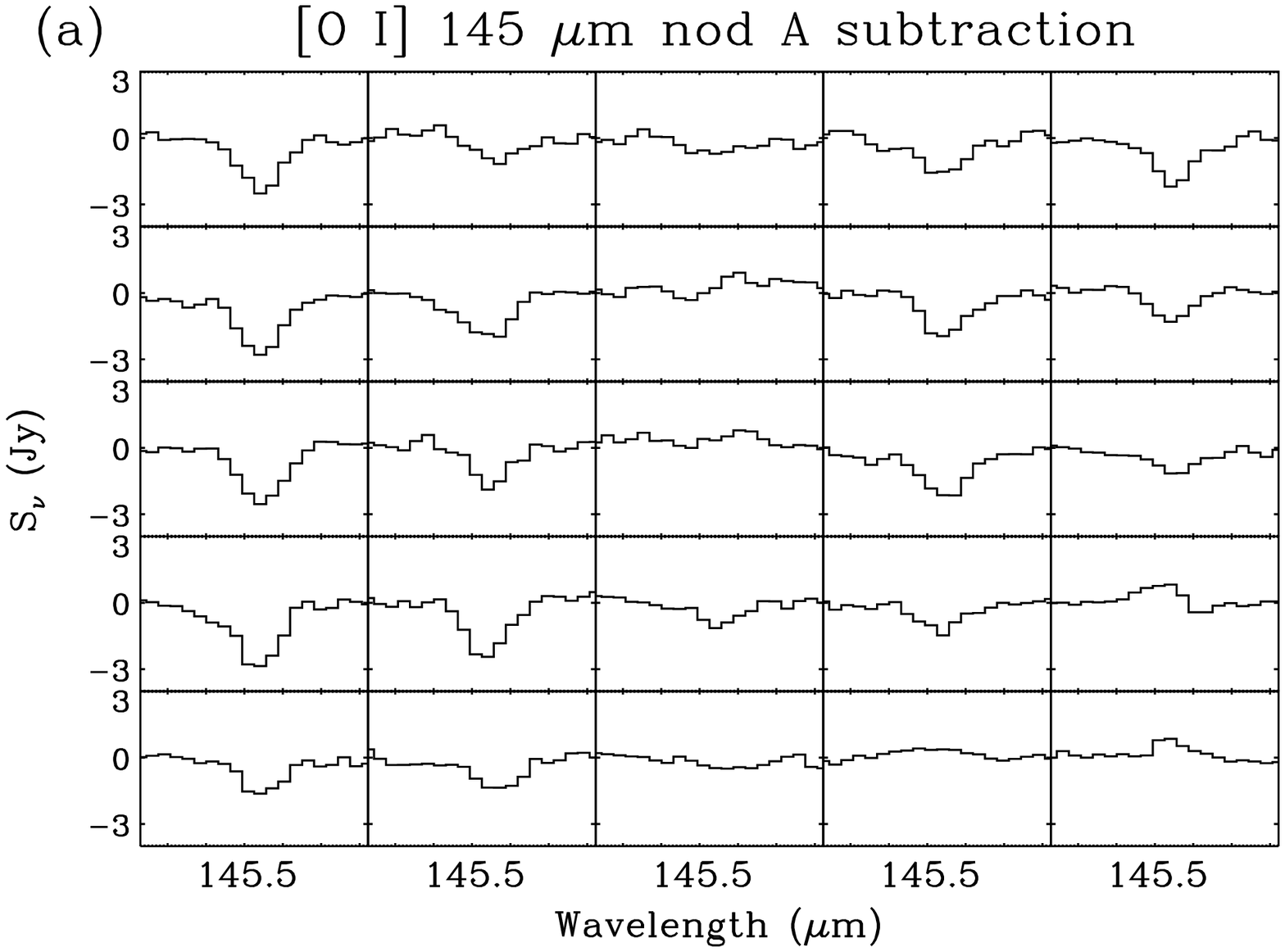}
 \includegraphics[width=0.5 \textwidth]{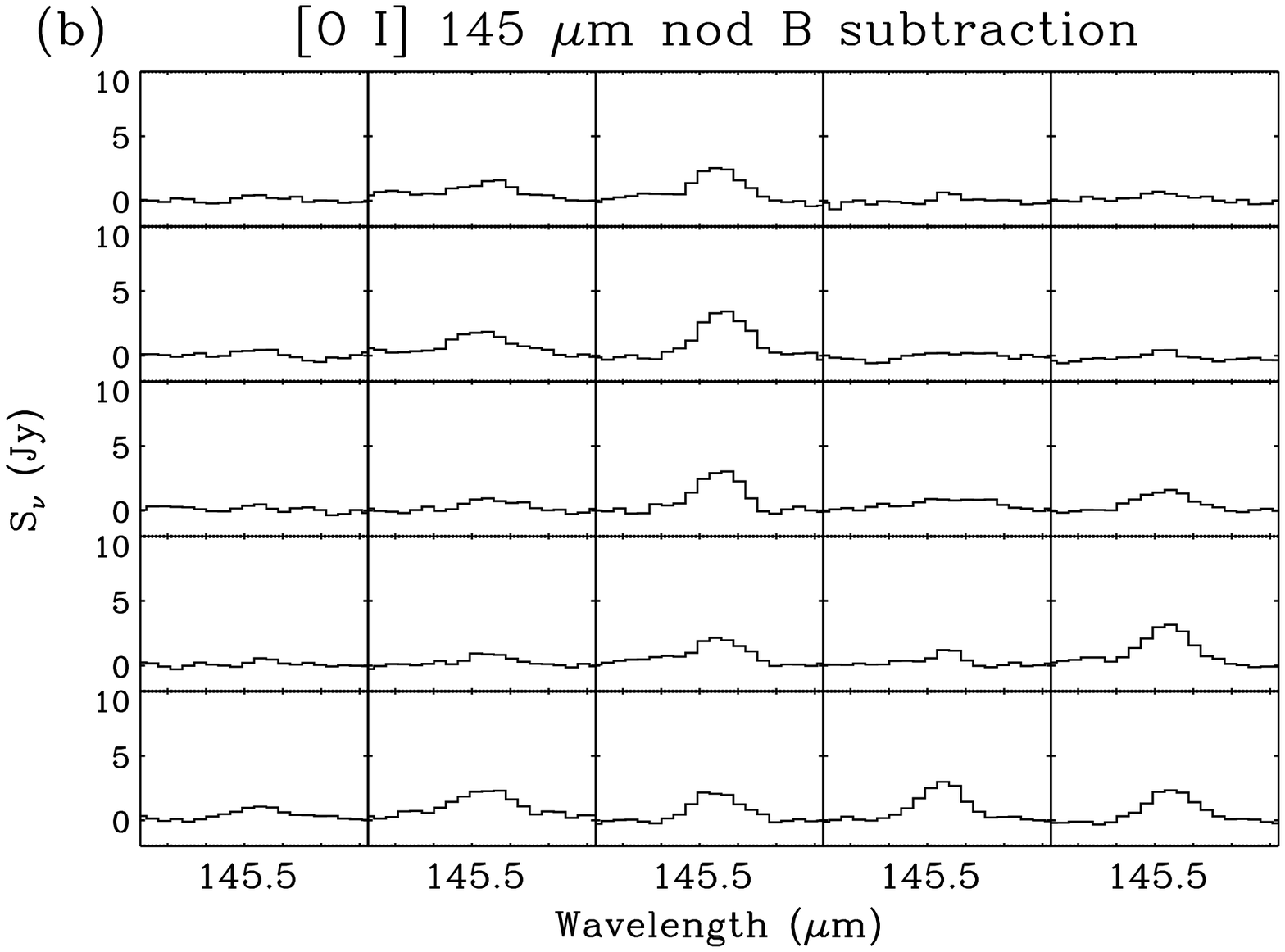} 
 \caption{The same as Figure~\ref{linemap_nod_OI63} but for [O I] 145 $\mu$m line.}
 \label{linemap_nod_OI145} 
\end{figure*}

\begin{figure*}
 \includegraphics[width=0.5 \textwidth]{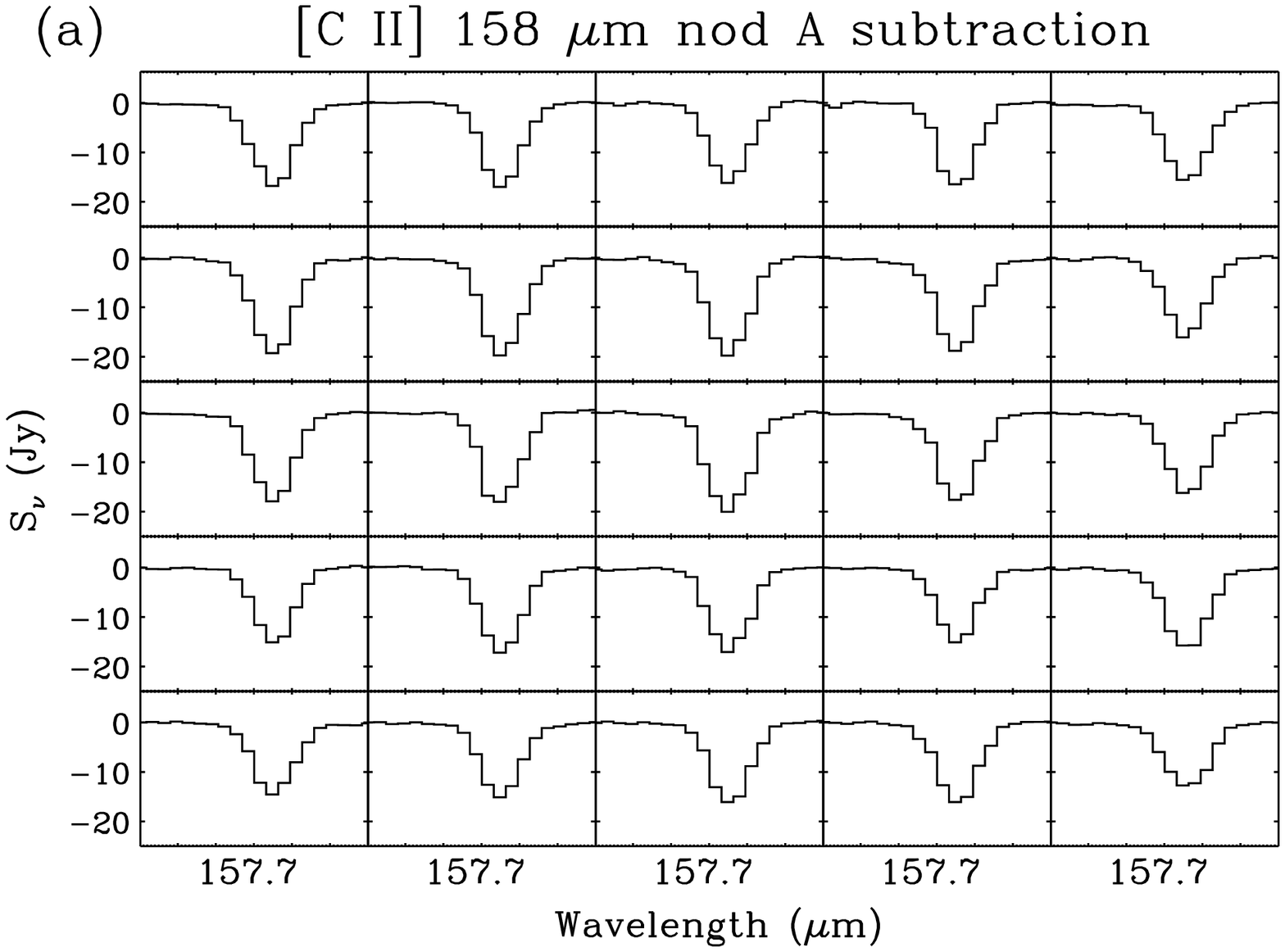}
 \includegraphics[width=0.5 \textwidth]{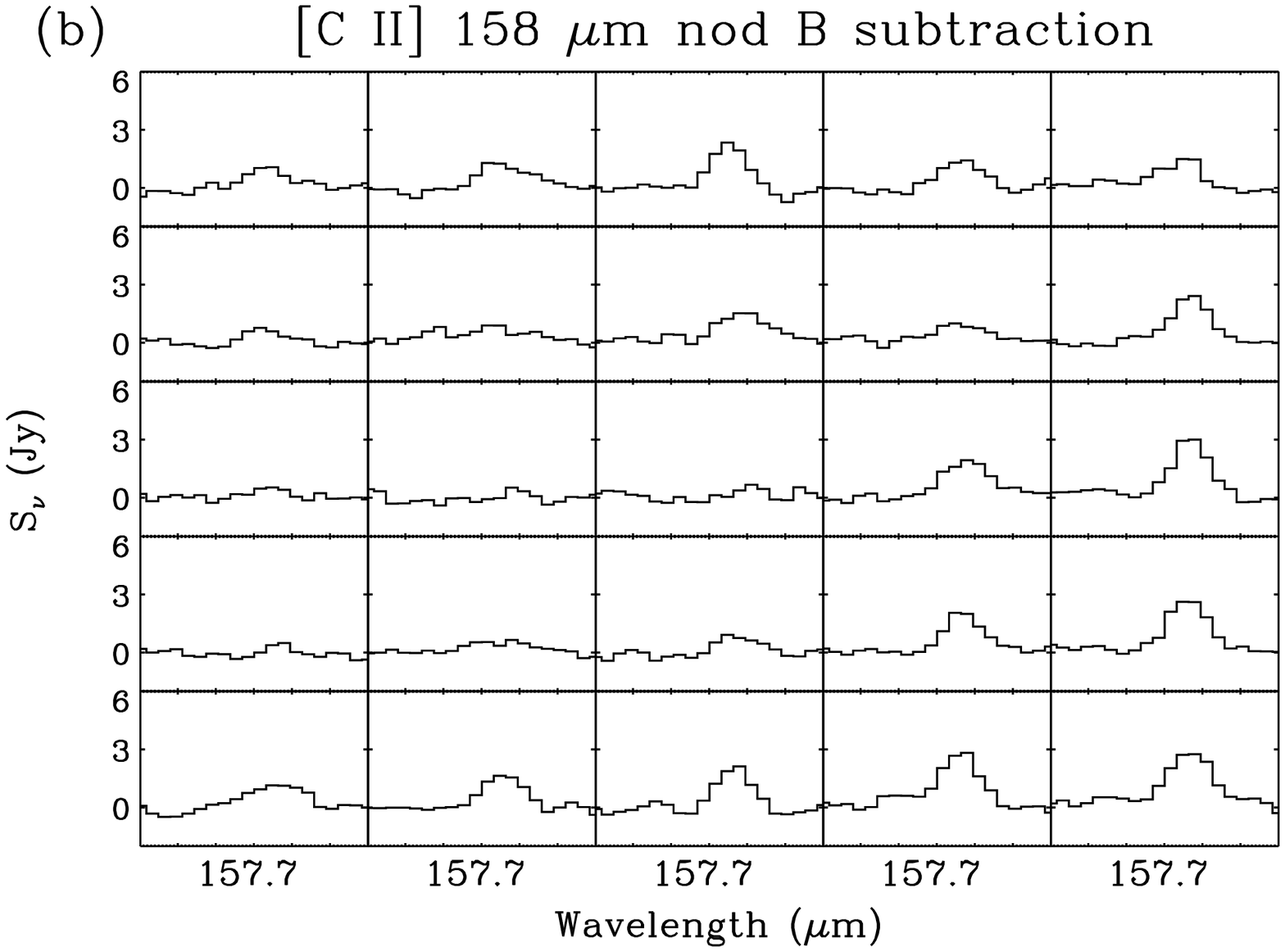} 
 \caption{The same as Figure~\ref{linemap_nod_OI63} and~\ref{linemap_nod_OI145} but for [C II] 158 $\mu$m line. The [C II] emission line is detected at only one nod observation. The other nod position was contaminated by the strong [C II] emission, resulting in the deep absorption feature in the source spectrum as seen in (a).}
 \label{linemap_nod_CII} 
\end{figure*}

\begin{figure*}
\includegraphics[scale=0.6]{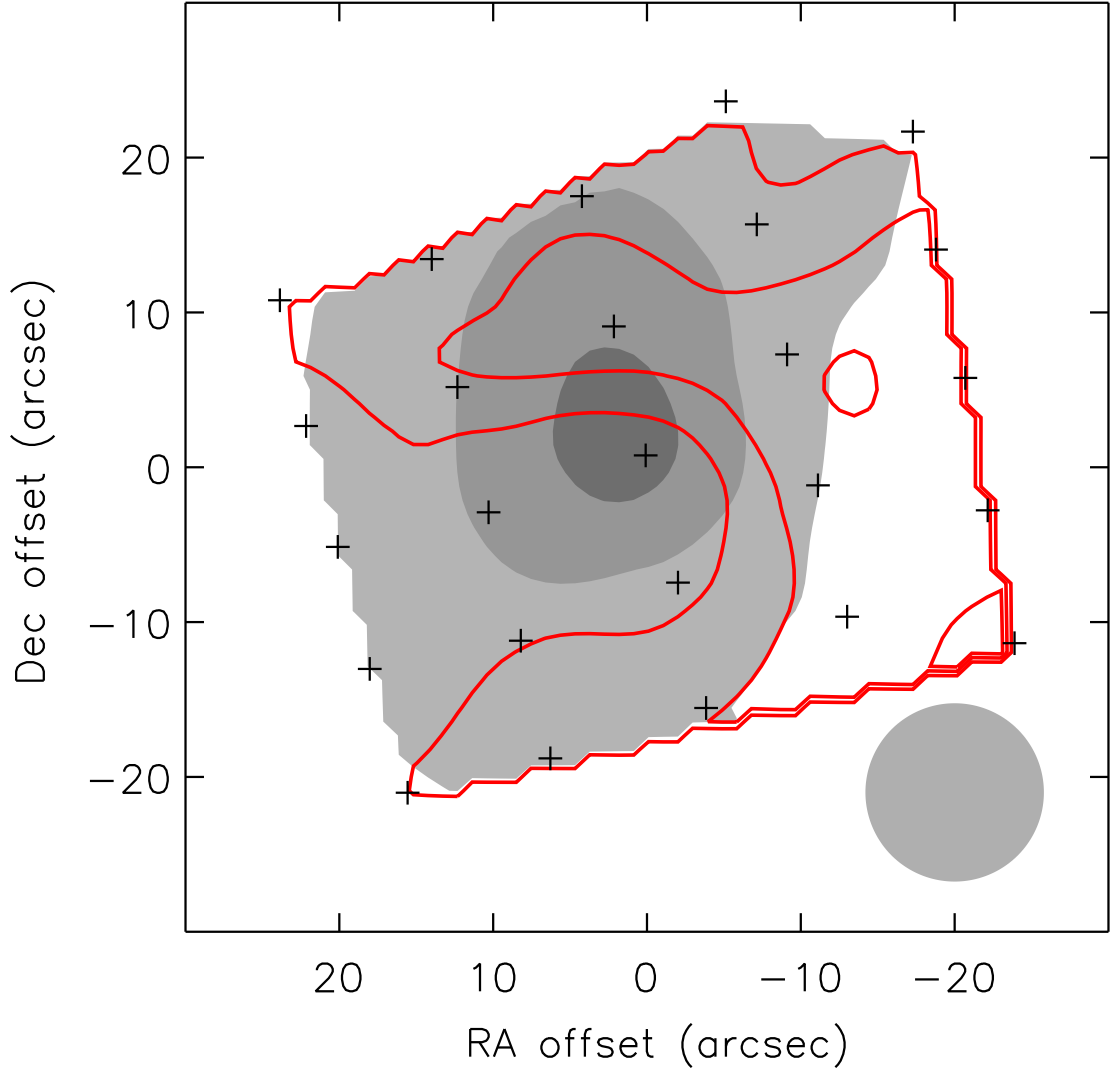}
 \centering
 \caption{Contour map of [C II] 158 $\mu$m line flux (red solid line) on top of the local continuum emission at 151 $\mu$m (gray scale). The line flux was calculated from Figure~\ref{linemap_nod_CII}b. The positions of PACS pixels are indicated as the black crosses. The gray filled circle at the right bottom shows the PACS beam size at 158 $\mu$m. Contour levels and gray scales are increasing in 30, 60, and 90 \% of the peak flux. } 
 \label{contour_CII} 
\end{figure*}

 Figure~\ref{linemap_nod_OI63} to Figure~\ref{linemap_nod_CII} show the atomic line maps where the spectrum of each nod position was subtracted separately. The [O I] line at 63 $\mu$m is clearly seen as an emission line in both subtractions and its morphology is similar to that found in the final data reduction (Figure~\ref{linemap_OI63}), suggesting that it is less affected by emission in reference positions. In maps of the [O I] 145 $\mu$m and [C II] 158 $\mu$m (Figure~\ref{linemap_nod_OI145} and Figure~\ref{linemap_nod_CII}), however, the left panel subtractions show an absorption feature because of the emission in the nod A position itself. However, the other subtractions in the right panel, using nod B, show emission lines. Therefore, the relative distribution of the [O I] 145 $\mu$m and [C II] 158 $\mu$m emission using nod B could be on-source emission although the actual intensity would be stronger if [O I] 145 $\mu$m and [C II] 158 $\mu$m emission exists in the nod B position. The [C II] intensity contour map (Figure~\ref{contour_CII}) uses the nod B subtraction in Figure~\ref{linemap_nod_CII}b. The [C II] emission does not peak in the central spaxel and it has no spatial correlation with [O I] emission
 (Figure~\ref{contour_line_conti}b and e), indicating that they are not excited by the same mechanism as [O I].

 Although there is a possibility that the atomic lines are contaminated by the nod positions, in this analysis we use the atomic lines of the nod B subtraction assuming the result is representative of the source.

\section{LINE ANALYSIS} \label{sec:line}
 Recently, a number of studies have presented the results of rotational diagrams and non-LTE analysis of molecular line emission detected with \textit{Herschel}/PACS \citep{Neufeld12, Green13, Karska13, Lee13, Manoj13, Lee14}. \citet{Green13} has described the simple rotational diagram of all DIGIT sources with molecular lines including GSS30-IRS1; GSS30-IRS1 is generally consistent with other embedded sources (Green et al. in prep.). 
 
 As clearly seen in Sect.~\ref{sec:result}, there is a different trend between atomic and molecular lines; molecular lines are centered on GSS30-IRS1 and are more compact compared to atomic lines. In the Taurus sources, [C II] emission is too weak to see the spatial distribution in most sources, except for L1551-IRS1 and TMC1 \citep{Lee14}. The [C II] emission in those two sources is well-correlated with [O I] emission, related to the outflow direction. However, the spatial distributions of [O I] and [C II] lines in GSS30-IRS1 have no correlation with each other. We provide the results of the rotational diagram and non-LTE analysis of molecular emission without much detailed description because the results are not very different from previous work toward other embedded YSOs. In this paper, we will focus more on the analysis of atomic lines.

 \subsection{Molecular Lines}
 \subsubsection*{Rotational Diagrams}
 In the rotational diagram, CO ladders are fitted by two rotational temperatures with a breakpoint at \Eup{} = 1800 K as done in \citet{Green13} and \citet{Lee14}. We refer to those papers for the fundamental concept of rotational diagrams. For the ``warm'' CO component, the rotational temperature is \Trot{} $\simeq$ 359 $\pm$ 4 K, and the ``hot'' component is fitted by \Trot{} $\simeq$ 783 $\pm$ 32 K. However, the apparent rotational temperatures of H$_{2}$O (\Trot{} $\simeq$ 197 $\pm$ 2 K for ortho-H$_{2}$O, and \Trot{} $\simeq$ 171 $\pm$ 5 K for para-H$_{2}$O) and OH (\Trot{} $\simeq$ 193 $\pm$ 7 K for $^{2}{\Pi}_{3/2}$ ladders, and \Trot{} $\simeq$ 128 $\pm$ 5 K for $^{2}{\Pi}_{1/2}$ ladders) are much lower than that of CO. H$_{2}$O and OH are usually sub-thermally excited due to their high critical densities. The derived rotational temperatures and total number of each molecules agree well with other DIGIT embedded sources. All results of rotational diagrams are found in Table~\ref{tbl_rot}.

 \subsubsection*{Non-LTE Analysis}
 For a more realistic analysis of molecular excitation, we utilized the RADEX escape probability code \citep{van der Tak07} to understand the physical conditions surrounding GSS30-IRS1 assuming a plane-parallel molecular cloud. We adopted the same ranges of physical parameters ($T_\textrm{kin}$, $n(\rm H_2$), and $N$(molecule)/$\Delta\textit{v}$) as used in \citet{Lee14} to find the best-fit models. The details of the modeling procedure with RADEX can be found in \citet{Lee13} and \citet{Lee14}. According to the non-LTE LVG models, the CO emission lines favor the highest temperature ($T_\textrm{kin}$ = 5000 K), and a power-law temperature distribution can fit the CO emission well, indicating the sub-thermal solution of CO. Our results for CO show a similar trend to those found in TMC1 and TMR1 \citep{Lee14}. For H$_{2}$O, one component model explains the observed H$_{2}$O emission better than the power-law model since the power-law model is meaningful only when the lines are optically thin. The best-fit model temperature (500 K) in the one component model for H$_{2}$O is much lower than that of CO or that of H$_{2}$O for all Taurus sources \citep{Lee14}. For OH, \citet{Lee13} reported that the IR-pumping plays a role only for the highest energy level (\Eup{} = 875) transition. However, our best-fit model suggests the model without IR-pumping fits better the highest energy transition than the model with IR-pumping. The model parameters of our best-fit model of each molecule are listed in Table~\ref{tbl_radex}.

 \subsection{Atomic Lines}
 \subsubsection*{The Origin of the [C II] emission}\label{sec:CII}
  The Oph molecular cloud is a good region to study the [C II] emission, which is known as a tracer of the PDRs excited by stellar or interstellar radiation field: (1) the cloud contains three B-type stars, (2) is close to us, and (3) is at a high enough galactic latitude to avoid the emission from the galactic plane \citep{Yui93}. Our PACS observations also show extended [C II] emission around GSS30-IRS1 (Figure~\ref{contour_CII}). According to \citet{Giannini01}, the detection of [C II] lines originating from embedded low luminosity YSOs is very unlikely due to their low luminosity and high density of envelope. Therefore, the detection of the [C II] line without any correlation of its spatial distribution with any other molecular emission indicates an external PDR activity around GSS30-IRS1.    
 
 \citet{Yui93} and \citet{Liseau99} reported three possible candidates of external UV sources: HD 147889 (B2 V), S 1 (B3 V), and SR 3 (B7 V) in the Oph cloud. Their visual extinctions are \textit{A}$_\textrm{v}$ = 4.6, 12.5, and 6.2 mag, respectively \citep{Elias78}. Both studies assume that HD 147889, which is the hottest among the three stars, is the dominant external UV source, though S 1 is closer in projection to the Oph cloud. Based on the model by \citet{kaufman99}, we interpret the strength of the far-ultraviolet (FUV) radiation field, \G0 (in units of the Habing Field, 1.6 $\times$ 10$^{-3}$ erg cm$^{-2}$ s$^{-1}$) combined with density of H nuclei (\textit{n}) and [C II] 158 $\mu$m intensity. 
 
 As we measured the [C II] line fluxes from the spaxels where [C II] line is detected in emission (Figure~\ref{linemap_nod_CII}b), we can calculate the total intensity of [C II] lines using emitting area with an assumption that the sky emission at the nod B position is insignificant. 

   The intensity of [C II] calculated over the PACS FOV is 1.1 $\times$ 10$^{-5}$ erg cm$^{-2}$ s$^{-1}$ sr$^{-1}$. If we adopt the density of 10$^{4}$ $\leq$ $\textit{n}$(H) $\leq$ 10$^{5}$ at the radius greater than 2000 AU, which is calculated from our dust continuum modeling (see Sect.~\ref{sec:continuum}), the strength of \G0 is in the range of 3 to 20 based on the PDR model from \citet{kaufman99}. Liseau's empirical model, based on their observed angular distribution of the [C II] line fluxes, also reveals that the value of \G0 in the Oph cloud is in the range of 10 to 100. The derived \G0 value gives us the lower-limit because there might be the [C II] emission in nod B position while we assume no emission at the nod B position.

 \subsubsection*{The Origin of the [O I] emission}
  Although not as significantly as the [C II] emission, the [O I] lines at 63 and 145 $\mu$m are also affected by the sky emission as seen in Figure~\ref{linemap_nod_OI63}a and Figure~\ref{linemap_nod_OI145}a. In particular, no emission line at 145 $\mu$m is seen at the central position of Figure~\ref{linemap_nod_OI145}a, indicating that the sky emission from the nod A position is comparable to the source flux. In order to calculate the actual intensity of [O I], we applied the same method to [O I] lines as done for [C II] except that the [O I] lines are measured from the PACS 15 spaxels (red crosses in Figure~\ref{Spitzer_PACS}). The [O I] intensity is 2.2 $\times$ 10$^{-4}$ and 1.2 $\times$ 10$^{-5}$ erg cm$^{-2}$ s$^{-1}$ sr$^{-1}$ for 63 and 145 $\mu$m, respectively. The derived [O I] line ratio between 145 $\mu$m and 63 $\mu$m suggests that the strength of \G0 can reach up to 10$^{6}$ \citep{kaufman99}, which is unrealistically high compared to other studies and our [C II] analysis.
  
  The [O I] emission seems associated with the HCO$^{+}$ flow \citep{Jorgensen09}, which is extended beyond the PACS FOV. It is also peaked at the central position both at 63 and 145 $\mu$m (Figure~\ref{contour_line_conti}b and e). This line morphology suggests that the [O I] emission is more directly related to the source itself, unlike the [C II] line. Therefore, a fast dissociative shock, which is produced by strongly collimated jets, or an irradiated C-shock along the outflow cavity wall, which is exposed by the UV photons from the central source, may cause most of the [O I] emission in GSS30-IRS1. 
  
\begin{figure*}
 \includegraphics[width=1 \textwidth]{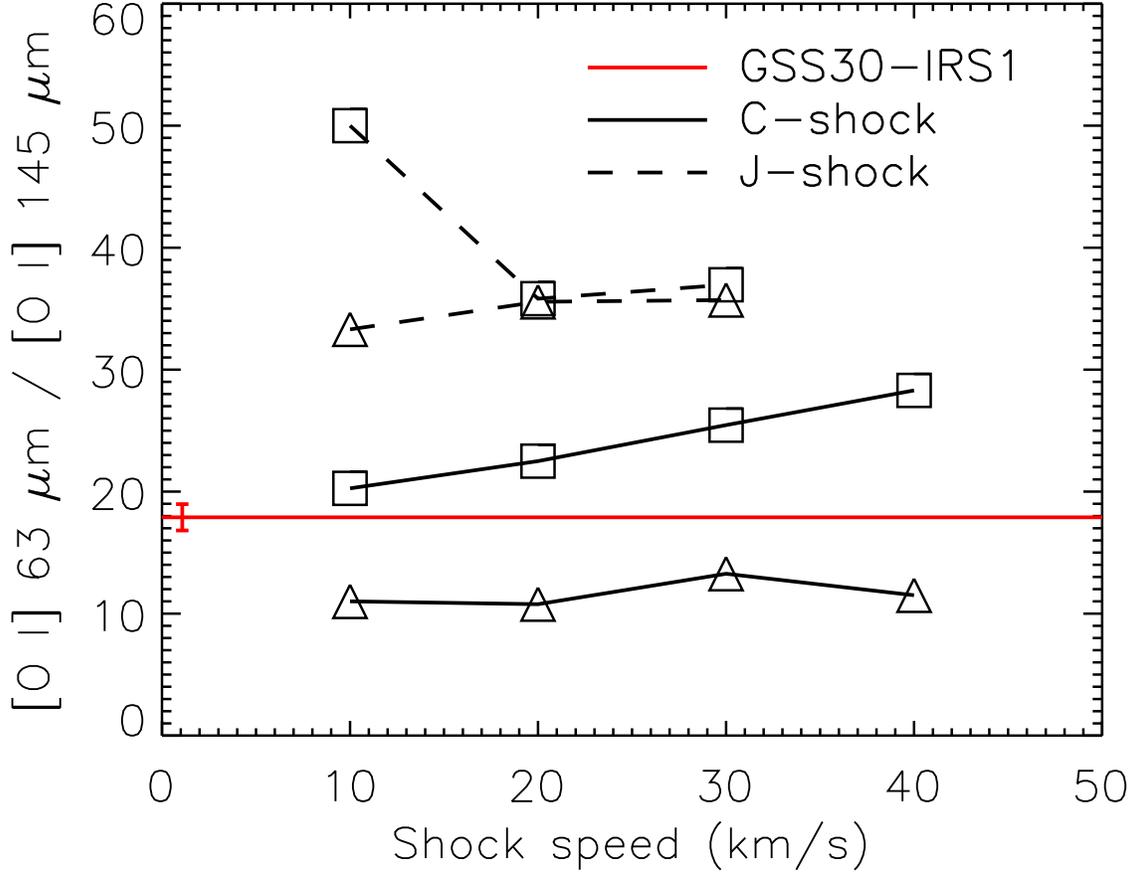}
 \centering
 \caption{The line ratio of [O I] 63 $\mu$m and 145 $\mu$m as a function of shock condition. Triangles are for $n(\rm H_2)= 2\times10^{4}$ cm$^{-3}$ and squares are for $n(\rm H_2)=2\times 10^{5}$ cm $^{-3}$. Black solid and dashed lines indicate C- and J-shock models, respectively. Red colors represent the flux ratio in GSS30-IRS1 and the vertical lines represent the uncertainties. }
 \label{shock_ratio} 
\end{figure*}

\begin{figure*}
 \includegraphics[width=1 \textwidth]{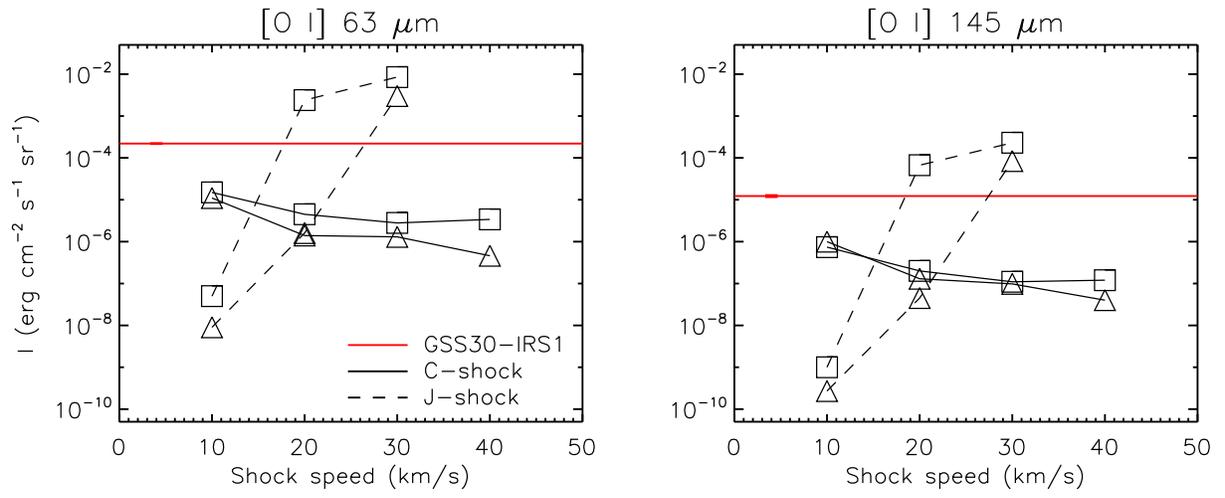}
 \centering
 \caption{The same as Figure~\ref{shock_ratio} but for the average intensity of [O I] 63 $\mu$m and 145 $\mu$m lines. }
 \label{shock_intensity} 
\end{figure*}

  The interactions between the outflow/jet and the dense envelope produce highly energetic shocks which propagate into molecular gas, consisting of H$_{2}$, CO, and H$_{2}$O. The abundances of those species strongly depend on the shock condition because the shocks could dissociate the molecules. Therefore relative intensities of those gas emission lines can characterize the type and speed of associated shocks. Thus, we compare the observed line fluxes with the shock models by \citet{Flower10}, who presented models for C- and J-type shocks, where the shock speed, $\textit{v}_\textrm{s}$, ranges from 10 km s$^{-1}$ to 40 km s$^{-1}$, and pre-shock densities, $\textit{n}_\textrm{H}$, from 2 $\times$ 10$^{4}$ cm$^{-3}$ to 2 $\times$ 10$^{5}$ cm$^{-3}$. The shock speed of 40 km s$^{-1}$ is only for the C-shock. According to the [O I] 63/145 $\mu$m flux ratio in GSS30-IRS1, the [O I] flux ratio is well fitted by a high density C-shock model (Figure~\ref{shock_ratio}). We also compared the actual intensity of [O I], which is calculated above, as done in \citet{Flower10}. Both [O I] lines are matched by a high density J-shock model with \textit{v} = 17 -- 27 km s$^{-1}$ (Figure~\ref{shock_intensity}). The [O I] 63 $\mu$m line is excited more easily than the [O I] 145 $\mu$m line due to its lower excitation energy (228 K). Therefore, if the J-shocks actually play an important role in GSS30-IRS1 as expected by the actual intensities of the [O I] lines, the low [O I] flux ratio, which is better explained by C-shocks, suggests that the [O I] 63 $\mu$m line is optically thick.

 \subsection{Line Luminosities}
  In Table~\ref{tbl_lum}, we present the FIR continuum and total line luminosities in the wavelength range covered by PACS as well as the FIR line luminosities of each species. The definition of total FIR line luminosity, $\textit{L}_\textrm{FIR}$, was adopted from \citet{Nisini02}: $\textit{L}_\textrm{FIR}$ = $\textit{L}_\textrm{CO}$ + $\textit{L}_\textrm{H$_{2}$O}$ + $\textit{L}_\textrm{OH}$ + $\textit{L}_\textrm{[O I]}$. In the PACS wavelength range, fractional contribution of the total line luminosity to the total FIR continuum luminosity is $\sim 1.9 \times 10^{-3}$, indicating that the cooling by the dust is dominant in FIR radiation. 
  
  According to \citet{Giannini01} and \citet{Nisini02}, the ratio of the total molecular cooling to the bolometric luminosity ($\textit{L}_\textrm{mol}$/\Lbol) in the FIR wavelengths is significantly higher ($\textit{L}_\textrm{mol}$/\Lbol{} $>$ 1 $\times$ 10$^{-3}$) for Class 0 sources than for Class I sources ($\textit{L}_\textrm{mol}$/\Lbol{} $<$ 10$^{-3}$), suggesting that the outflow power is greater in Class 0 than Class I. In GSS30-IRS1, the ratio of the FIR molecular cooling in the PACS range to the bolometric luminosity is $\sim$ 8 $\times$ 10$^{-4}$, which is consistent with the Class I range. 
  
  The percentage contributions to total line cooling are 21, 36, 11, and 32 $\%$ for CO, H$_{2}$O, OH, and [O I], respectively, in GSS30-IRS1. Thus the cooling through molecular emission is dominant (68 $\%$ to $\textit{L}_\textrm{FIR}$), but [O I] is also a significant coolant (32 $\%$). By contrast, analysis of the Class I protostar L1551-IRS5 indicates [O I] contributes around 70 $\%$ to the total line cooling, with almost no contribution of H$_{2}$O \citep{Lee14}. The 36 $\%$ contributed cooling from H$_{2}$O, similar to [O I], in GSS30 is indicative of the lower dissociation rate of H$_{2}$O.
  
\begin{figure*}
 \includegraphics[width=1 \textwidth]{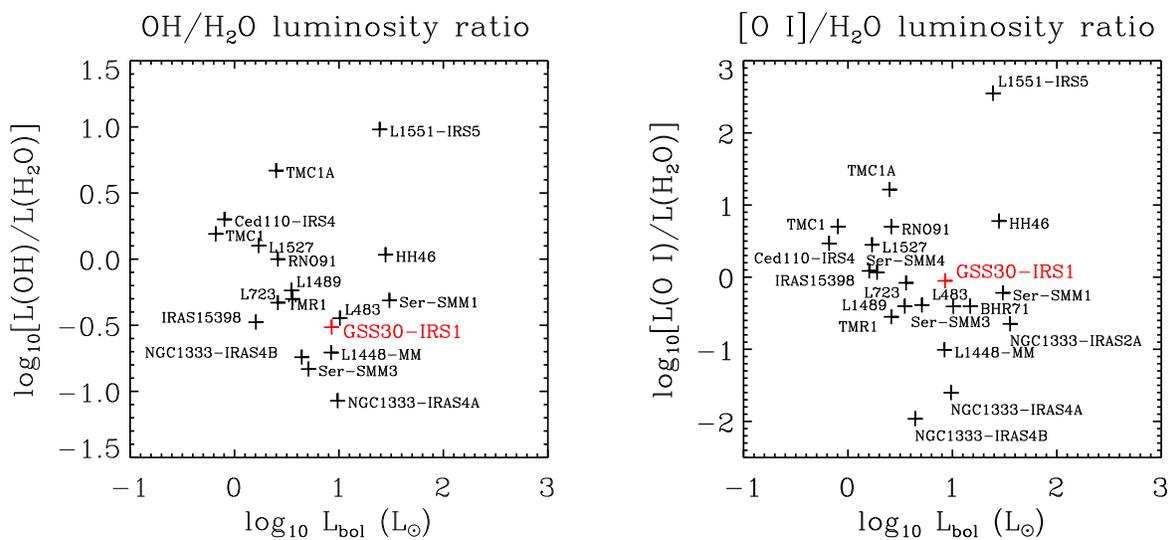}
 \centering
 \caption{The total line luminosity ratios toward the embedded sources. \textit{Left}: the ratio between $\textit{L}_\textrm{OH}$ and $\textit{L}_\textrm{H$_{2}$O}$. \textit{Right}: the ratio of $\textit{L}_\textrm{[O I]}$ and $\textit{L}_\textrm{H$_{2}$O}$. GSS30-IRS1 is presented as the red crosses.}
 \label{linelum_ratio} 
\end{figure*}

  If GSS30-IRS1 is associated with a PDR excited by the UV photons from a nearby luminous B type star or accretion to central protostar, the abundances of [O I] and OH would be enhanced compared to that of H$_{2}$O \citep{Hollenbach97} due to the dissociation of molecules. We thus calculated the total line luminosity ratios and compared the ratios with Taurus sources \citep{Lee14}, L1448-MM \citep{Lee13}, and all WISH sources \citep{Karska13}. We find that GSS30-IRS1 has a small OH/H$_{2}$O luminosity ratio compared to other embedded sources (Figure~\ref{linelum_ratio}). This does not necessarily imply that H$_{2}$O is not being dissociated, since the OH, the transient species in the dissociation of H$_{2}$O, survives for a very short time before its dissociation to H and O \citep{van Disoeck83}. The [O I]/H$_{2}$O ratio of GSS30-IRS1 is close to unity, indicative of the possible influence of high energy photons \citep{Lee14}.

 \section{DUST RADIATIVE TRANSFER MODELS} \label{sec:continuum}
 \subsubsection*{Internal Heating}  \label{sec:internal}
 \citet{Kristensen12} found the envelope physical structure of GSS30-IRS1 by fitting its SED with the 1-D continuum radiative transfer code DUSTY \citep{Ivezic97}. They considered a density profile, \textit{n} $\propto$ $\textit{r}^{-\textit{p}}$, where \textit{n} is the density, \textit{r} is the radius, and \textit{p} is the power-law index. The fitted SED data covered from 24 $\mu$m (\textit{Spitzer}-MIPS) to 850 $\mu$m (SCUBA) including the \textit{Herschel}/PACS observations. However, the NIR and MIR datapoints cannot be fit by this 1-D envelope, instead requiring a disk structure, which radiates predominantly at shorter wavelengths \citep{Kristensen12}. In addition to the disk, embedded YSOs have strong outflows to develop bipolar cavities. Thus, we construct a 2-D SED model using the deconvolved spectrum of IRS1 assuming that IRS1 is the dominant internal heating source in the vicinity of GSS30.
  
 To explore the disk and envelope structures of GSS30-IRS1, we modeled the SED using the Monte Carlo radiative transfer package RADMC-3D v3.1 \citep{Dullemond04}. The main inputs are the density structure of the dusty circumstellar material, the dust opacity tables, and the stellar parameters. The model assumes that the scattering is isotropic. We adopted the same dust grain models used by \citet{Robitaille06}, a mixture of silicates and graphite grains \citep{Laor93}. We used different grain models for different regions in the disk and the envelope as follows: ``Disk midplane'' dust grain model described in Table 3 of \citet{Whitney03} for the dense region ($n_\textrm{H$_{2}$}$ $>$ 10$^{10}$ cm$^{-3}$) and the ``KMH'' dust grain model for the rest of the circumstellar geometry \citep{Kim94}. The grain model for the disk midplane has a size distribution which decays exponentially for grain sizes between 50 $\mu$m and 1 mm. The grains in the envelope are slightly larger than ISM grains. For the detailed grain properties, see \citet{Whitney03} and \citet{Robitaille06}. 
    
 To determine initial parameters, we used the online SED fitting tool\footnote{See http://www.astro.wisc.edu/protostars} by \citet{Robitaille06}. This tool offers a grid of 200,000 SEDs including stars, disks, and envelopes with a wide range of physical parameters. The SED is sensitive to the total internal luminosity and the inner density of the envelope, especially in FIR range. The total internal luminosity is the sum of the stellar luminosity and accretion luminosity, where the stellar luminosity is given by $\textit{L}_{\ast}$ $\propto$ $\textit{R}_{\ast}^{2} \textit{T}_{\ast}^{2}$ and accretion luminosity, $\textit{L}_{acc}$, is a function of stellar parameters and disk accretion rate $\dot{\textit{M}}_\textrm{disk}$. Since there is high degeneracy between accretion and stellar luminosity, we simplify the fitting by adjusting the total luminosity only via stellar luminosity. The envelope density is parameterized in terms of the envelope accretion rate. We further explored those parameters using a fixed distance of 120 pc to find the best-fit model based on the initial model parameters obtained from the online tool (Robitaille's model ID : 3005892). In our modeling, fixed model parameters among those derived from Robitaille's online fitting tool are listed in Table~\ref{tbl_robitaille}. 
      
 We construct the density structures of the three components consisting of disk, envelope, and outflow cavity, following \citet{Robitaille06}.
   
 The density structure for the envelope can be described as 
   \begin{eqnarray}
  \rho(r,\theta) = \frac{\dot{M}_\textrm{env}}{4\pi(GM_{\ast}R_\textrm{c}^{3})^{1/2}} \Big(\frac{r}{R_\textit{c}}\Big)^{-3/2}
   \Big(1+\frac{\mu}{\mu_{0}}\Big)^{-1/2} \Big(\frac{\mu}{\mu_{0}} + \frac{2\mu_{0}^{2}R_\textrm{c}}{r}\Big)^{-1},
\end{eqnarray}    \label{den_envelope}
 where $\dot{\textit{M}}_\textrm{env}$ is the envelope accretion rate, $\textit{M}_{\ast}$ is the stellar mass, and $\textit{R}_\textrm{c}$ is the centrifugal radius which is usually associated with an outer radius of the disk. This centrifugal radius determines the flattening of envelope structure. $\mu$ is cos $\theta$, and $\mu_\textit{0}$ is cosine of the angle of a streamline of infalling particles as r $\to$ $\infty$ \citep{Ulrich76, Terebey84}. It can be solved by the equation of parabolic motion:
   \begin{equation} \label{mu0}
   \mu_{0}^{3} + \mu_{0}\big(\frac{r}{R_{c}} - 1\big) - \mu\big(\frac{r}{R_{c}}\big) = 0.
   \end{equation}
      
   The disk density profile is assumed to be flared with an accretion motion characterized by \citet{Shakura73}, \citet{Lynden-Bell74}, and \citet{Hartmann98}, 
   \begin{eqnarray} \label{den_disk}
   \rho(\varpi,z) = \rho_{0}\Big(1-\sqrt{\frac{R_{\ast}}{\varpi}}\Big)
   \Big(\frac{R_{\ast}}{\varpi} \Big)^{\alpha}\exp\Big\{-\frac{1}{2}\Big[\frac{z}{h}\Big]^{2}\Big\},
   \end{eqnarray}
    
 where $\textit{R}_{\ast}$ is the stellar radius, and $\rho_{0}$ is the normalization constant calculated from the integral of the density over the whole disk that corresponds to $\textit{M}_\textrm{disk}$. $\varpi$ and $\mathnormal{z}$ are the radius and height in the cylindrical coordinates. The disk scale height, \textit{h}, is defined as \textit{h} $\propto$ $\varpi$ $^\beta$, where $\beta$ is the disk flaring factor. The radial density exponent, $\alpha$, is obtained from $\alpha$ = $\beta$ + 1. 
   
 We also include the outflow cavity in the model described by 
   \begin{equation} \label{den_outflow}
   z_{\textrm{cav}} = c \varpi^{1.5}.
   \end{equation}   
 \textit{c} is $\textit{R}_\textrm{e,max}$/($\textit{R}_\textrm{e,max}$ $\tan\theta_\textrm{cav}$)$^{1.5}$, where $\textit{R}_\textrm{e,max}$ is the outer radius of the envelope, and $\theta_\textrm{cav}$ is the cavity opening angle.   
   
 We compared the model to the photometry in Figure~\ref{SED_whole}. We also obtained the binned IRS and PACS fluxes measured at 9, 12, 18, 25, 70, 120, and 160 $\mu$m from IRS and PACS spectra for more reliable fits. The best-fit is determined by the following logarithmic deviation, where   
   \begin{eqnarray} \label{chi}
   R = \sum_{i=1}^{\it N}[\omega_{i} \vert \rm ln(\it F_{\lambda_{i},\rm o}/\it F_{\lambda_{i},\rm m})\vert ]/ \it N.
   \end{eqnarray}
 $F_\textrm{$\lambda_{i}$,o}$ is the observed fluxes at each wavelength, $F_\textrm{$\lambda_{i}$,m}$ is the model fluxes at each wavelength, $N$ is the number of data points, and $\omega_{i}$ is the inverse of the fractional uncertainty in each point \citep{Fischer12}.  
    
\begin{figure*}
 \includegraphics[width=1 \textwidth]{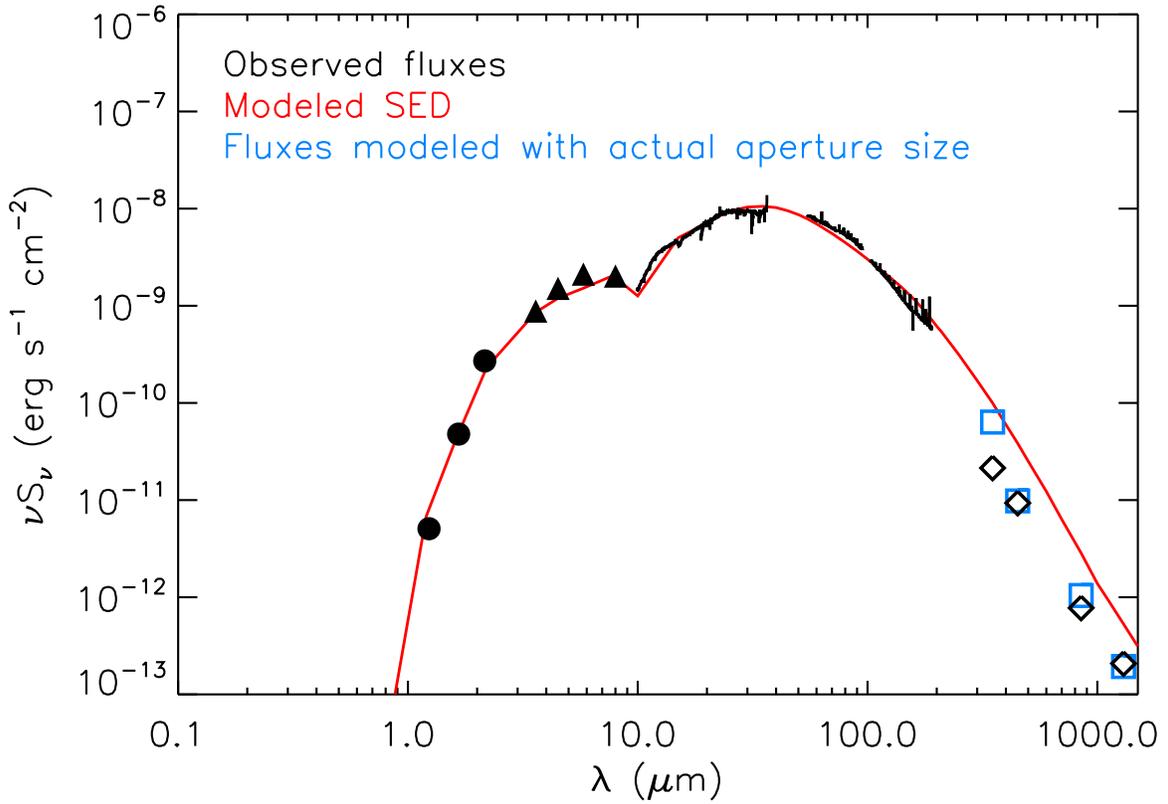}
 \caption{The best-fit SED model (red solid line) for the \textit{Herschel}/PACS deconvolved spectrum of IRS1. Data symbols are the same as in Figure~\ref{SED_whole}. Blue squares represent the model fluxes with the actual aperture sizes.}
 \label{SED_internal} 
\end{figure*}

 In Figure~\ref{SED_internal}, we present the best-fit model for the deconvolved SED of IRS1. It has \textit{R} = 3.1 (models with \textit{R} $<$ 4 provide acceptable fits within observational uncertainty). We also projected the sub-mm fluxes at observed aperture sizes. The model overestimate of the flux compared with the data from SHARC-II 350 $\mu$m is acceptable because the observation was insensitive to extended emission such as GSS30-IRS1 \citep{Wu07}. The model has a total luminosity of $\textit{L}_\textrm{tot}$ = 10 $\pm$ 1 \Lsun{}, an envelope mass of $\textit{M}_\textrm{env}$ = 0.23 \Msun, and an envelope accretion rate of $\dot{\textit{M}}_\textrm{env}$ = 3 $\times$ 10$^{-6}$ \Msun{} $yr^{-1}$. The total luminosity derived from the best-fit model is greater than the observed \Lbol{} of 8.5 \Lsun{}, indicating that sub-mm data does not trace the total extent of GSS30-IRS1. The envelope mass falls within the mass derived from sub-mm fluxes \citep{Enoch09}. The envelope accretion rate derived from the best-fit model is the same as that found in Robitaille's online fitting tool. The cavity opening angle is 5.2 $^{\circ}$ and the inclination angle is 41 $^{\circ}$, where 0 $^{\circ}$ is face-on. The interstellar extinction at V band is 12.19, which is consistent with the mean extinctions derived from \citet{Evans09}. When compared to the 1D envelope model \citep{Kristensen12}, this SED model much better describes the NIR fluxes from the disks.
  
\begin{figure*}
 \includegraphics[width=1 \textwidth]{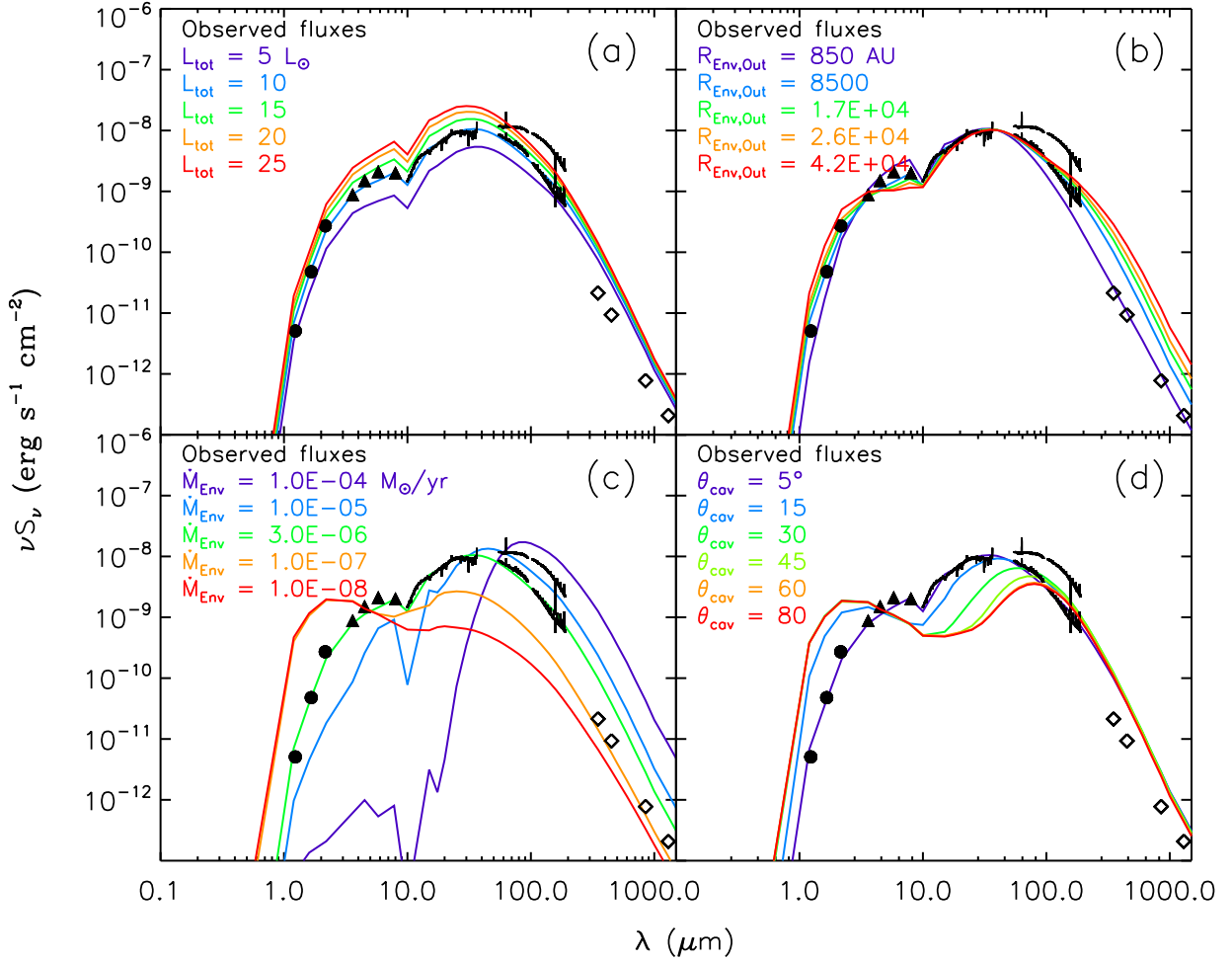}
 \caption{Exploration of parameter space: (a) total luminosities, (b) outer radius of envelope, (c) envelope mass infall rate, and (d) outflow cavity angle. The different line colors in each box represent the SEDs calculated with different values of a given parameter. No adjustment of these parameters can fit the total PACS fluxes after subtracting the deconvolved spectra of IRS2 and IRS3.}
 \label{SED_internal_test} 
\end{figure*}

 After modeling the deconvolved IRS1 spectrum, we now model the PACS total emission over 5 $\times$ 5 spaxels after subtracting deconvolved emission from IRS2 and IRS3 (hereafter PACS total fluxes) in order to study whether the PACS total fluxes can be explained solely by the internal heating by adjusting the model parameters of the best-fit model of IRS1. In order to fit the PACS total fluxes (Figure~\ref{SED_whole}, blue lines), models need to produce higher fluxes at FIR wavelengths and a smaller slope at longer wavelengths compared to the model fitting of the deconvolved SED of IRS1. When we increase the total internal luminosity in the model, the flux in FIR rises, but the fluxes at NIR are also enhanced (Figure~\ref{SED_internal_test}a). If the outer radius of the envelope increases, the slope of SED at long wavelength decreases (Figure~\ref{SED_internal_test}b). With higher envelope mass infall rates that provide high opacity, the NIR flux decreases because the envelope blocks the photons from the disk (Figure~\ref{SED_internal_test}c). (Here, the parameter of ``mass infall rate" in the model really controls the opacity of the envelope.) The opening angle of the outflow cavity affects the SED only at a wavelength shorter than about 100 $\mu$m; fluxes at $\lambda$ $<$ 3 $\mu$m increase with opening angle, while fluxes at 3 $\mu$m $<$ $\lambda$ $<$ 100 $\mu$m decrease with opening angle (Figure~\ref{SED_internal_test}d).
  
 None of these models are a good fit to the total FIR derived from PACS, suggesting that other input parameters are needed. We also tried to fit the full SED using Robitaille's online SED fitting tool in order to confirm the above results because Robitaille's model deals with only the internal heating. The minimum chi-square model fails to fit the observed fluxes beyond 100 $\mu$m wavelengths. We note that before the outflow cavity is carved out, the envelope density structure of our model is assumed to be spherically symmetric in spite of two other YSOs located closely to IRS1; three YSOs are possibly gravitationally interacting to deviate the density structure from spherical symmetry. 
Geometrical effects, such as aspherical distributions may play a role, but the 70 $\mu$m emission from internal heating always arises on small scales. Only external heating can provide the extended excess emission that we observe.

 \subsubsection*{External Heating}  \label{sec:external}
 The possibility of an enhanced radiation field in the Oph cloud compared to the standard radiation field in the solar neighborhood has been previously studied in depth by \citet{Yui93} and \citet{Liseau99}. They suggested that the Oph region might be filled with strong UV radiation on a large scale, indicative of diffuse PDRs over the entire Oph region. \citet{Yui93} found that the UV radiation extends a few parsecs from the nearby B type star over the whole cloud. \citet{Lindberg12} reported that low-mass protostars cannot heat the surrounding envelope up to 30 K, but an external radiation field can explain the high temperature of 40 K that we found in the residual spectrum. In addition, our deconvolution results show that the fluxes measured over the PACS 25 spaxels greatly exceed the sum of each deconvolved flux of three YSOs, suggesting that external heating may contribute to the total flux density of GSS30-IRS1. Therefore, we discuss how the external radiation field has an effect on the PACS total IRS1 fluxes.
  
 The modeling of the SED with both internal and external heating was done with the same physical parameters obtained from the model with only internal heating. We also modeled the PACS fluxes in their observed apertures, because the PACS FOV cannot trace the outer region of the source where the external heating is more dominant than internal heating. 
  
\begin{figure*}
 \includegraphics[width=1 \textwidth]{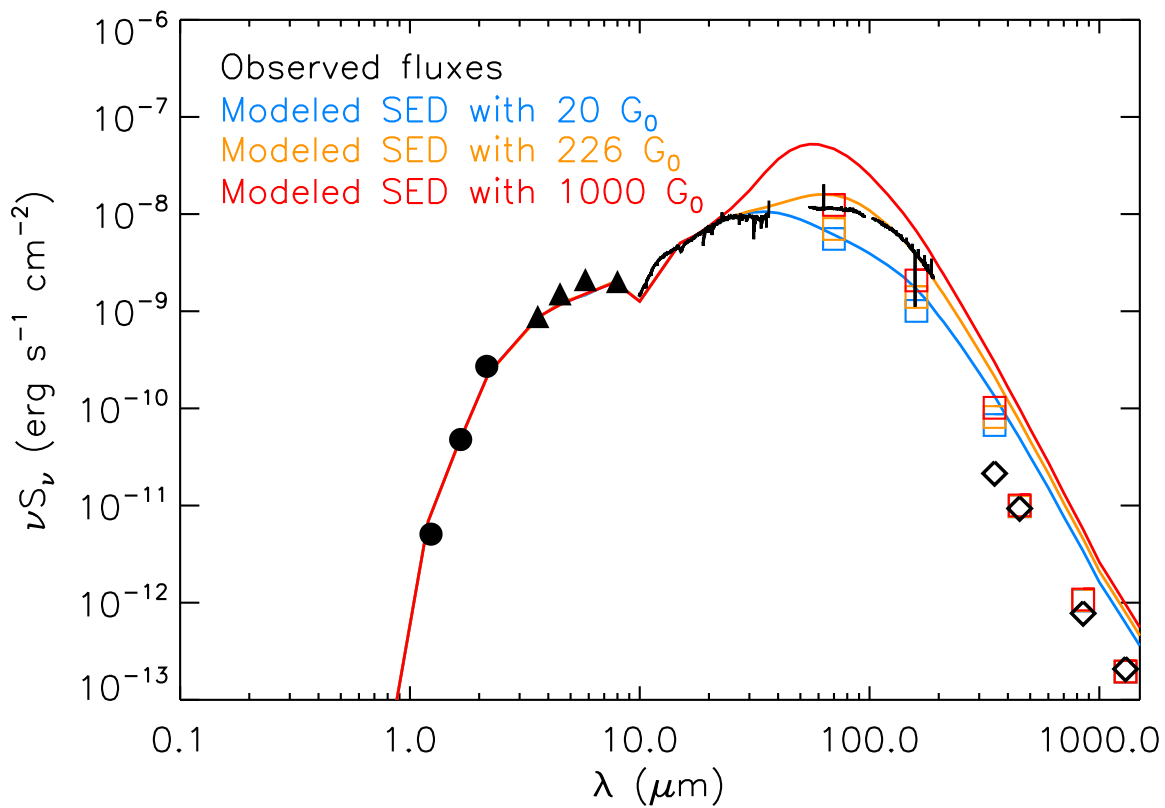}
 \caption{Three different best-fit models (blue, orange, and red solid lines) for the total fluxes after subtracting the deconvolved spectra of IRS2 and IRS3. The symbols for the observational data are the same as in Figure~\ref{SED_whole} except for the \textit{Herschel}/PACS. Squares represent the fluxes modeled with actual aperture sizes.}
 \label{SED_external_test} 
\end{figure*}

 First, we included an external radiation field of 20 \G0, expected to be the minimum \G0 from [C II] emission (Sect.~\ref{sec:CII}). The best-fit SED model with 20 \G0 is presented in Figure~\ref{SED_external_test} (blue line). However, the model fails to fit the observed fluxes in the PACS apertures (blue squares). Furthermore, we adopted the simple blackbody spectrum of a B2 V star as an external radiation field, and we set its position using the projected distance between the B2 V star and GSS30-IRS1, approximately 0.48 pc given a distance of 120 pc to the Oph cloud. If the effective temperature of the B2 V star is taken to be $\simeq$ 22,000 K \citep{Liseau99}, the star increases the radiation field by a factor of 226 \G0 at a distance of 0.48 pc. In this model, the total fluxes of the best-fit model correspond to the observed fluxes, but the model still cannot reproduce the fluxes in the PACS apertures (orange squares in Figure~\ref{SED_external_test}). According to our models, the UV flux must be enhanced more than 1000 \G0 in order to fit the PACS data (red squares in Figure~\ref{SED_external_test}). If the GSS30-IRS1 and the B2 V star are not in the same plane of sky, the 226 \G0, which is calculated at the projected distance of 0.48 pc, would give us the maximum \G0. Therefore, the value of 1000 \G0 is unrealistic. This test suggests that if heating only by the high energy photons is considered, the B2 V star cannot be the heating source.
  
\begin{figure*}
 \includegraphics[width=1 \textwidth]{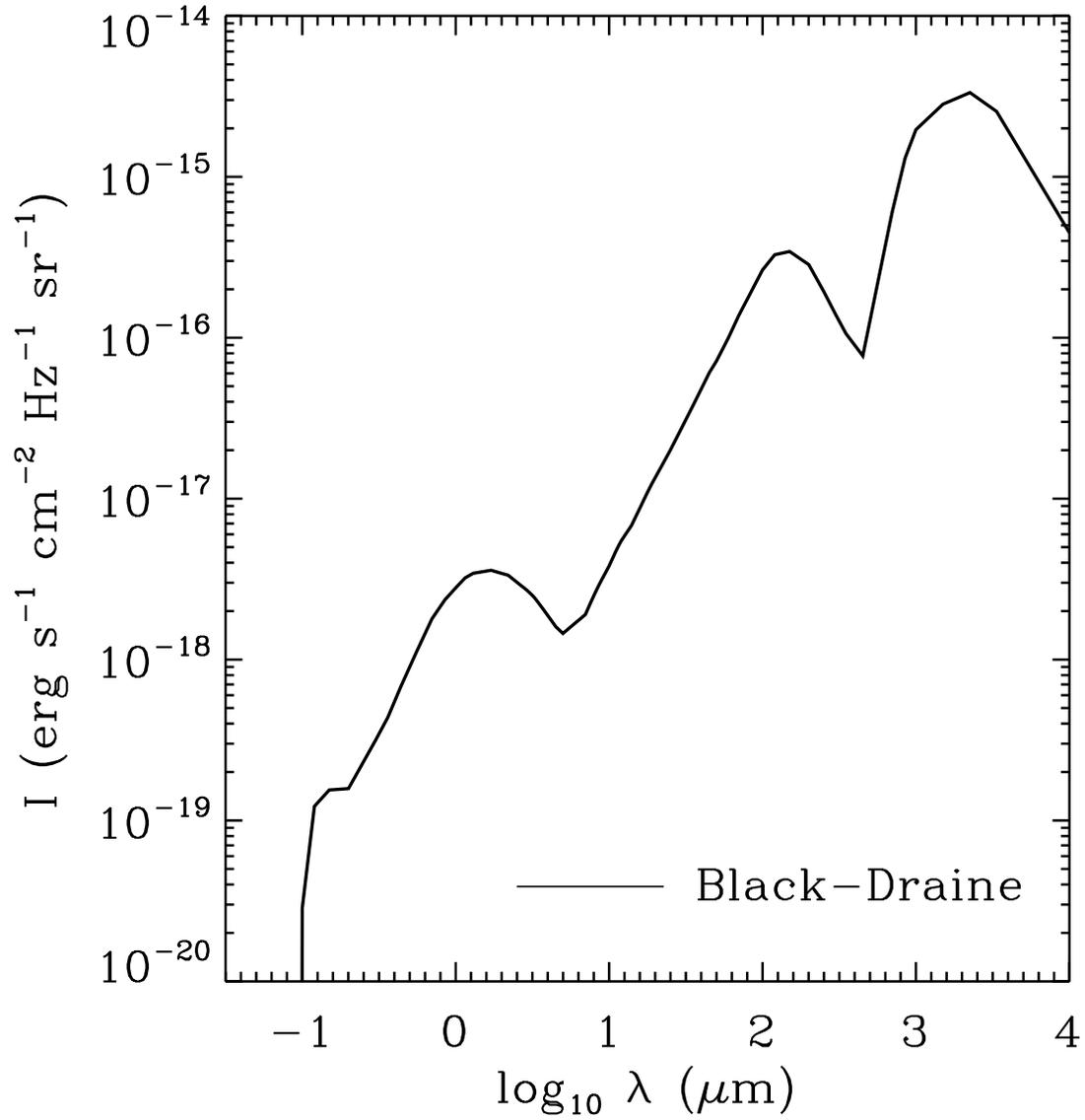}
 \caption{Plot of the standard interstellar radiation field used in the dust continuum modeling. It is the combination of \citet{Black94} (for $\lambda$ $\ge$ 0.36 $\mu$m) and \citet{Draine78} (for $\lambda$ $<$ 0.36 $\mu$m.)}
 \label{ISRF_BD} 
\end{figure*}

\begin{figure*}
 \includegraphics[width=1 \textwidth]{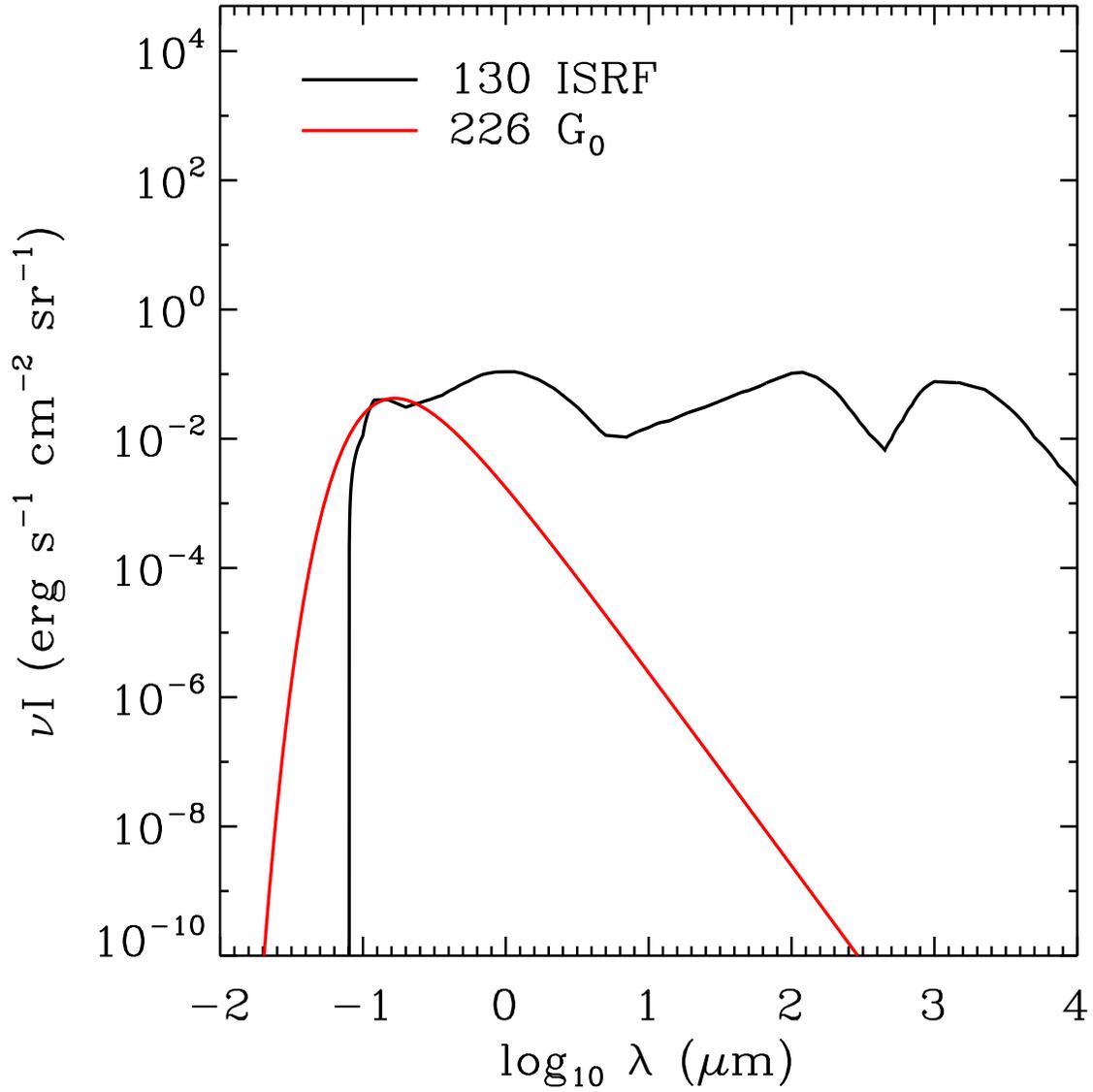}
 \caption{The spectrum of the interstellar radiation field enhanced by a factor of 130 (black). For comparison, we also present the radiation field of the B2 V star at the distance of 0.48 pc from the GSS30-IRS1 (red).}
 \label{ISRF_total} 
\end{figure*}

  Finally, we adopted the spectrum of ISRF from \citet{Evans01} to explore the external heating effect. They combined the spectra from \citet{Black94} for $\lambda$ $\ge$ 0.36 $\mu$m and \citet{Draine78} for $\lambda$ $<$ 0.36 $\mu$m (Figure~\ref{ISRF_BD}, the Black-Draine field). After modeling the SED with the ISRF enhanced by factors of 5 to 500, we find that the model with the ISRF enhanced by a factor of 130 can reproduce the observed PACS total fluxes well. If we calculate the \G0 value at the FUV range using this enhanced ISRF, it is 201, which is much smaller than the required enhancement (1000 \G0) when the spectrum of the B2 V star is adopted. Nevertheless, the ISRF enhanced by a factor of 130 can fit the observations because the ISRF spectrum has much higher intensities at wavelengths greater than the wavelength of intensity peak of the B2 V star as seen in Figure~\ref{ISRF_total}.
  
\begin{figure*}
 \includegraphics[width=1 \textwidth]{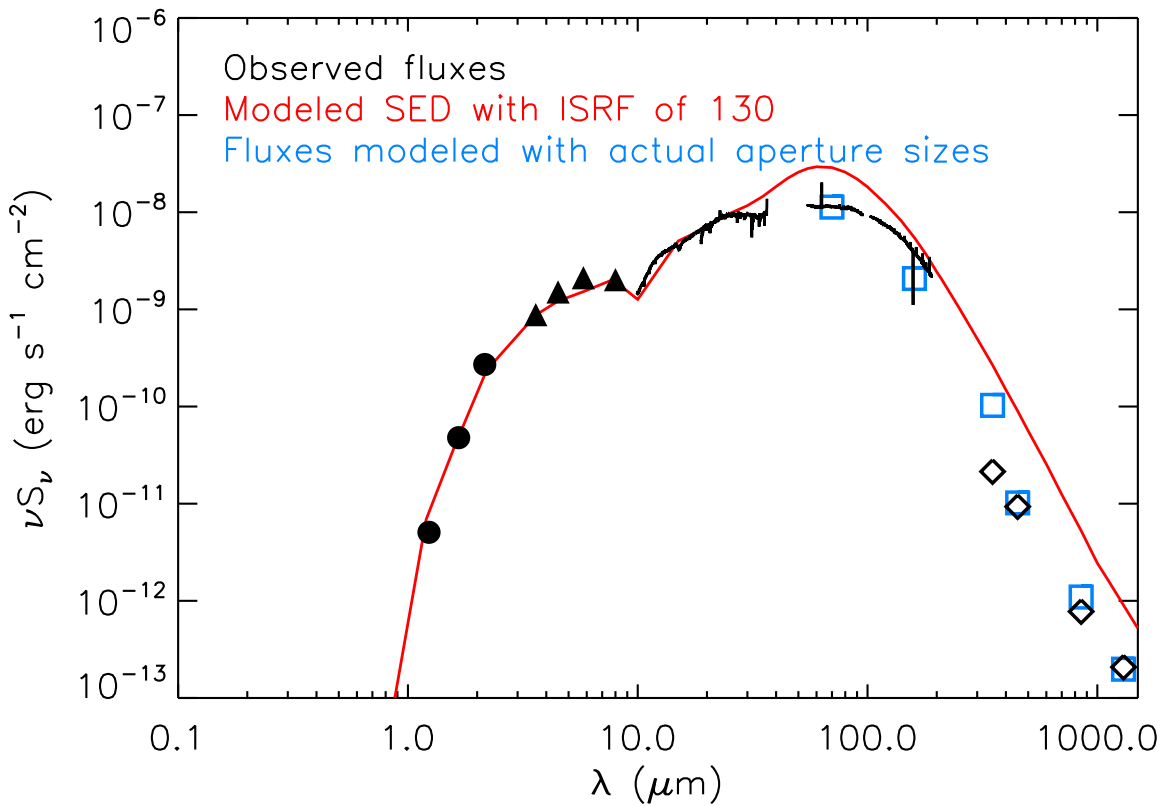}
 \caption{The best-fit SED model (red solid line) with the interstellar radiation field enhanced by a factor of 130 for the total \textit{Herschel}/PACS fluxes after subtracting the deconvolved fluxes by IRS2 and IRS3. The symbols for the observational data are the same as in Figure~\ref{SED_whole} except for the \textit{Herschel}/PACS. Blue squares represent the model fluxes with the actual aperture sizes. }
 \label{SED_external} 
\end{figure*}

  The best-fit SED model with the ISRF enhanced by a factor of 130 is presented in Figure~\ref{SED_external}. Our model photometry at 70 $\mu$m fits the observed total fluxes well, but there is a disagreement at 160 $\mu$m. It is possible that this flux deficit at 160 $\mu$m is caused by a larger structure surrounding GSS30-IRS1. In the SCUBA dust continuum map at 850 $\mu$m (Figure~\ref{Spitzer_PACS}b), the emission is extended beyond the PACS FOV and its peak is closer to IRS3 than IRS1. In addition, the 1.3 mm continuum flux density of IRS3 after background subtraction is greater than IRS1 by a factor of 2 \citep{Motte98}. In order to see how the core structure can influence the fluxes at the PACS longer wavelength, we increased the outer radius of the envelope by a factor of 2 compared to the best-fit model with a fixed constant density. Our simple test confirmed that cold dust material in the larger structure leads to the increase of flux at 160 $\mu$m, although the NIR flux decreased due to the added extinction. 
   
\begin{figure*}
 \includegraphics[width=1 \textwidth]{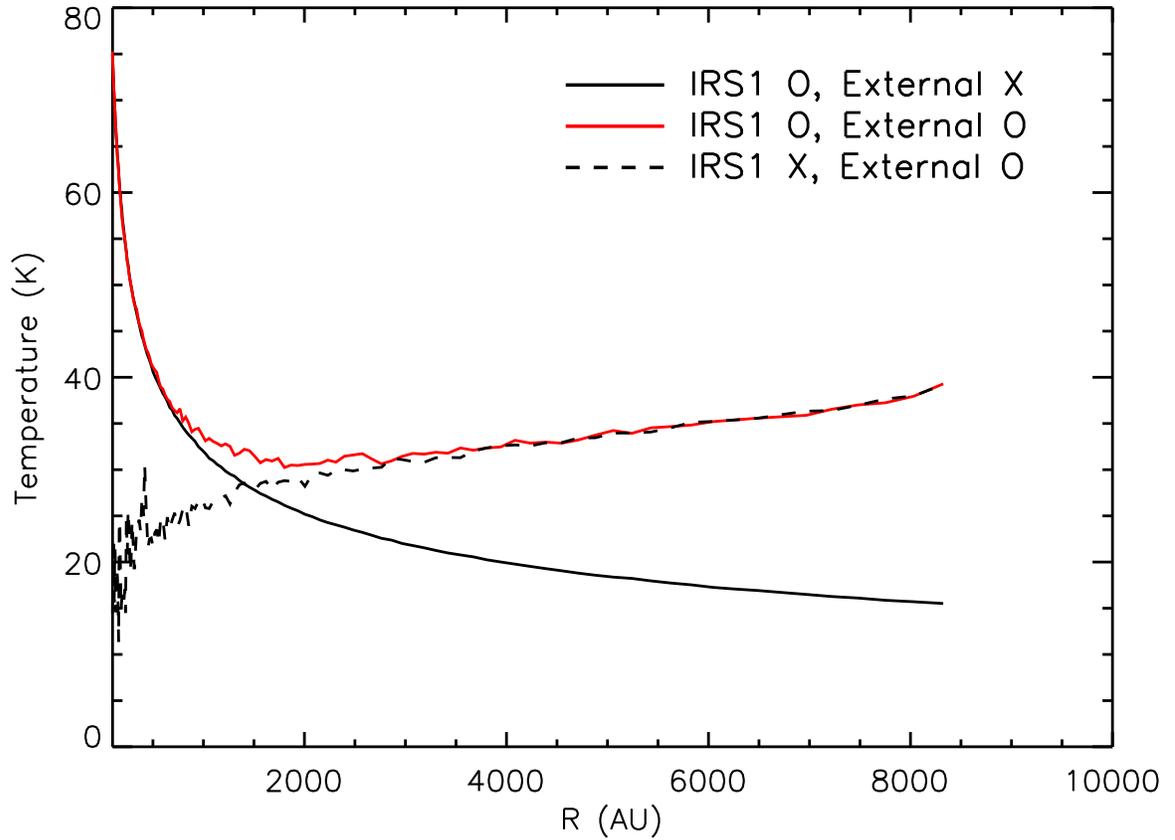}
 \caption{The dust temperature profile as a function of radius. The black solid line is the model including only he internal heating by IRS1. The same model but with the external heating by the enhanced interstellar radiation field is indicated as a red solid line. The black dashed line is represents the model without the central source but including the external radiation field. }
 \label{SED_external_temp} 
\end{figure*}

 The dust temperature profile derived from three different models is presented in Figure~\ref{SED_external_temp}. In the model without external heating, the temperature drops below 20 K beyond 4000 AU, indicating that internal heating cannot produce a temperature of 40 K, which we found from the residual spectrum, at large radii. However, the same model, with the interstellar radiation field enhanced by a factor of 130, can produce such high temperature at the outer envelope radius. Without a central heating source, the external heating model cannot explain the high temperature at the inner part of envelope ($\le$ 2000 AU).
 
\begin{figure*}
 \includegraphics[width=1 \textwidth]{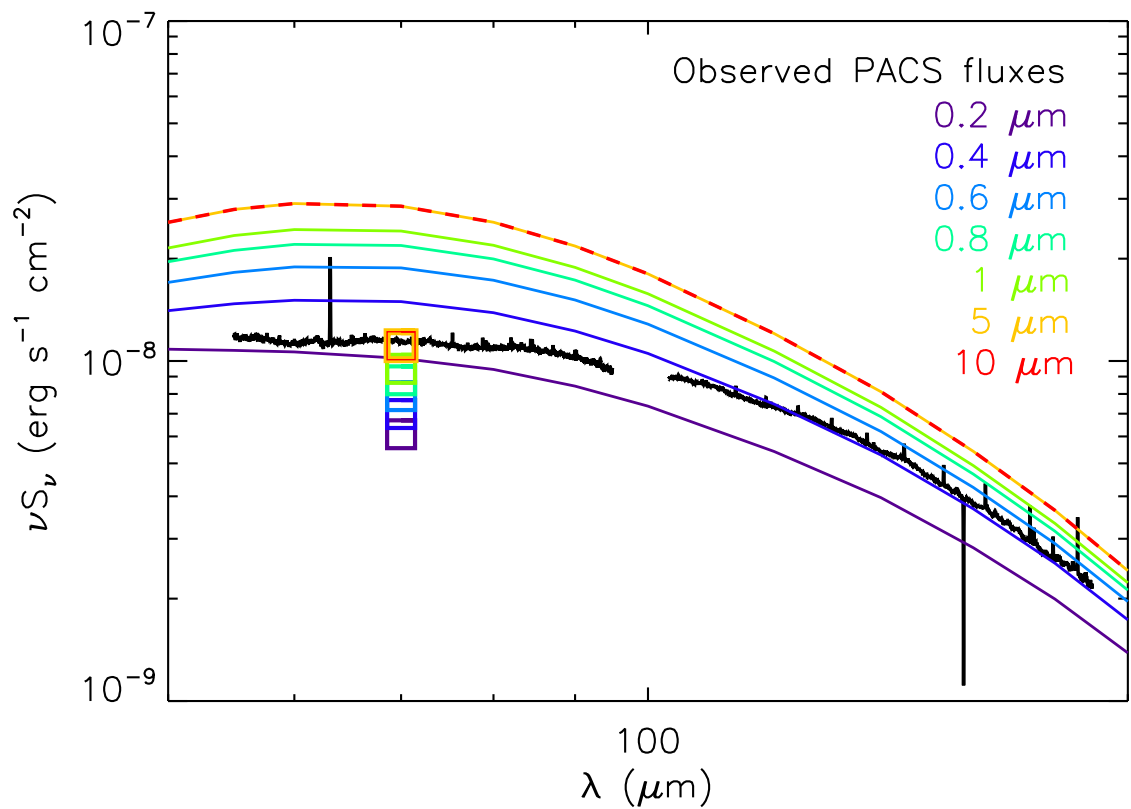}
 \caption{The SED models (colored solid and dashed lines) with the external heating by the ISRF for the total \textit{Herschel}/PACS fluxes after subtracting the deconvolved fluxes by IRS2 and IRS3 (black solid line). In order to test the contribution of low energy photons, we set the wavelength of the ISRF differently; the interstellar radiation field enhanced by a factor of 130 is used in the wavelength range from 0.01 $\mu$m to a certain wavelength, and the ISRF is set to be 0 in the rest part of wavelengths. Different colors represent different upper boundaries of wavelength of the ISRF, as marked in the upper right of the box. Squares represent the fluxes at 70 $\mu$m modeled with actual aperture sizes.}
 \label{SED_external_test_FIR} 
\end{figure*}

  Our various models show that dust continuum emission from the PACS 25 spaxels cannot be produced solely by the internal heating from the central object or UV fluxes from the B2 V star, but it can be explained by the enhanced Black-Draine interstellar radiation field, which has intensities increasing with wavelength, indicative of the importance of low energy photons for the dust heating. In order to test the contribution of low energy photons in our SED modeling, we enhance the interstellar radiation field only for the wavelength range from 0.01 $\mu$m to a certain boundary wavelength; except for this wavelength range, no radiation is assumed to exist. In Figure~\ref{SED_external_test_FIR}, each boundary of upper wavelength is presented with different colors. Our test supports that low energy photons (0.25 eV $<$ \textit{h}$\nu$ $<$ 6 eV; 0.2 $\mu$m $<$ $\lambda$ $<$ 5 $\mu$m) as well as the UV photons (6 eV $<$ \textit{h}$\nu$ $<$ 13.6 eV; 0.1 $\mu$m $<$ $\lambda$ $<$ 0.2 $\mu$m) play a role in heating dust material in the envelope. (However, the photons with energies lower than 0.25 eV ($\lambda$ $>$ 5 $\mu$m) do not affect the dust heating significantly as seen with the red dashed line in Figure~\ref{SED_external_test_FIR}.) 
Therefore, stars with spectral types later than B can contribute to the external heating of YSOs. On the other hand, the [C II] line emission is predominantly produced by UV photons from the B2 V star since carbon is ionized by UV photons, specifically.
 
 \citet{Green13} presented the ratios of flux densities extracted from the PACS 3 $\times$ 3 spaxels and the PACS central spaxel for the 15 well-centered sources including five sources (GSS30, VLA1623, Elias29, WL12, and IRS63) in the Oph cloud. According to their results, the ratios in the Oph sources (except for WL12 and IRS63) greatly exceed the PACS/PSF correction. Therefore, the continuum models of sources in the Oph cloud (or other externally heated sources) need more careful modeling to avoid misfitting the observed data, especially when their spatial resolutions are poor.

\section{SUMMARY} \label{sec:summary}
  We analyzed a Class I YSO, GSS30-IRS1, observed with the \textit{Herschel}/PACS, as a part of the DIGIT key program. We investigated its physical conditions based on the PACS observations by analysis of molecular and atomic lines and modeling the dust continuum emission using the 3D dust continuum radiative transfer code (RADMC-3D).

1. The three YSOs around GSS30 (IRS1, IRS2, and IRS3) are decomposed from the total FIR continuum flux across the full PACS array (5 $\times$ 5 spaxels). The deconvolved fluxes of IRS1 are consistent with the PACS central flux, corrected for the PACS/PSF. At longer PACS wavelengths, the deconvolved IRS1 flux is comparable to the deconvolved IRS3 flux.  

2. Almost 75 molecular and atomic lines are detected: the ladders of CO from \textit{J} = 14--13 to 40--39, 38 H$_{2}$O lines from 100 K to 1500 K of excitation energy, 14 transitions of OH lines in two different ladders, and two atomic [O I] lines at 63 and 145 $\mu$m. The [C II] 158 $\mu$m emission line, which is known as a tracer of the PDRs, is also detected.

3. All detected emission lines peak in the central spaxel and are spatially compact, except for [O I] and [C II]. The [O I] emission is extended along the NE-SW direction, which is similar to the outflow direction reported in previous studies. However, the [C II] emission shows no correlation with any other line emission. One special feature of GSS30-IRS1 is that the continuum is extended toward the northeast direction, beyond the PACS/PSF, indicative of anisotropic external heating.

4. The rotational diagrams and non-LTE LVG models indicate similar molecular gas properties to other YSOs observed with \textit{Herschel}, indicating that any external radiation field in GSS30 does not have a great effect on the excitation conditions.

5. The [C II] emission provides information on the FUV radiation field around GSS30 (3 $\sim$ 20 \G0). Although the [C II] line can be excited by the UV photons from the nearby B type star, dust continuum emission over all PACS spaxels cannot be explained solely by a FUV radiation field.  

6. Our best-fit model of the PACS SED supports the idea that the enhanced interstellar radiation field is significant in GSS30. In this model, the dust temperature at the outer envelope radius corresponds to the temperature found in the residual spectrum after the deconvolution method. Thus, both internal and external heating must be considered to better constrain the physical structures of the envelope. Advanced dust radiative transfer models are required to understand the complex envelope morphology around GSS30, as well as those around other sources exposed to significant external radiation fields.

\acknowledgments
Support for this work, part of the \textit{Herschel} Open Time Key Project
Program, was provided by NASA through an award
issued by the Jet Propulsion Laboratory, California Institute of
Technology.
J.-E. L. is very grateful to the department of Astronomy, University of Texas
at Austin for the hospitality
provided to her from August 2013 to July 2014.
This research was supported by the Basic
Science Research Program through the National Research Foundation
of Korea (NRF) funded by the Ministry of Education of
the Korean government (grant No. NRF-2012R1A1A2044689).
This work was also supported by the BK21 plus program through the National Research Foundation (NRF) funded by the Ministry of Education of Korea


\clearpage
\begin{deluxetable}{ccccc}
\tablewidth{0pt}
\tabletypesize{\footnotesize}
\scriptsize
\tablecaption{Detected Line Fluxes of GSS30-IRS1 \label{tbl_flux}}
\tablehead{
\colhead{Species} 		&\colhead{Transition}	 	&\colhead{\Eup}	      &\colhead{$\lambda$}      &\colhead{Flux $^{a}$} \\
\colhead{ } 		&\colhead{ }	 	&\colhead{(K)}	      &\colhead{($\mu$m)}      &\colhead{(10$^{-18}$ W m$^{-2}$)}  }
\startdata
CO&40-39&     4512.67 & 65.69 &          130 $\pm$ 27\\
 &37-36&      3871.69 & 70.91 &          73 $\pm$ 10\\
 &36-35&      3668.78 & 72.84 &          131 $\pm$ 21\\
 &35-34&      3471.27 & 74.89 &          92 $\pm$ 9\\
 &34-33&      3279.15 & 77.06 &          146 $\pm$ 13\\
 &33-32&      3092.45 & 79.36 &          152 $\pm$ 17\\
 &32-31&      2911.15 & 81.81 &          155 $\pm$ 21\\
 &31-30&      2735.28 & 84.41 &          348 $\pm$ 14\\
 &30-29&      2564.83 & 87.19 &          194 $\pm$ 19\\
 &29-28&      2399.82 & 90.16 &          188 $\pm$ 15\\
 &28-27&      2240.24 & 93.35 &          230 $\pm$ 13\\
 &25-24&      1794.23 & 104.44 &         196 $\pm$ 8\\
 &24-23&      1656.47 & 108.76 &         232 $\pm$ 14\\
 &23-22&      1524.19 & 113.46 &         661 $\pm$ 24\\
 &22-21&      1397.38 & 118.58 &         282 $\pm$ 8\\
 &21-20&      1276.05 & 124.19 &         306 $\pm$ 7\\
 &20-19&      1160.20 & 130.37 &         285 $\pm$ 6\\
 &19-18&      1049.84 & 137.20 &         279 $\pm$ 6\\
 &18-17&      944.970 & 144.78 &         290 $\pm$ 13\\
 &17-16&      845.590 & 153.27 &         317 $\pm$ 10\\
 &16-15&      751.720 & 162.81 &         327 $\pm$ 11 \\
 &15-14&      663.350 & 173.63 &         320 $\pm$ 11 \\
 &14-13&      580.490 & 186.00 &         416 $\pm$ 13\\
\\
OH & $\frac{1}{2}$,$\frac{9}{2}$-$\frac{1}{2}$,$\frac{7}{2}$ &    875.100& 55.89&         147 $\pm$ 22\\
 & &      875.100& 55.95&          126 $\pm$ 40\\
 & $\frac{3}{2}$,$\frac{9}{2}$-$\frac{3}{2}$,$\frac{7}{2}$ &      512.100& 65.13&         608 $\pm$ 35\\
 & &      510.900& 65.28&          204 $\pm$ 30\\
 &  $\frac{1}{2}$,$\frac{7}{2}$-$\frac{1}{2}$,$\frac{5}{2}$ &     617.600& 71.17&          97$\pm$ 6\\
 & &      617.900& 71.22&          134 $\pm$ 52\\
 &  $\frac{1}{2}$,$\frac{1}{2}$-$\frac{3}{2}$,$\frac{3}{2}$ &     181.900& 79.12&         200 $\pm$ 6\\
 & &      181.700& 79.18&          177 $\pm$ 12\\
 &  $\frac{3}{2}$,$\frac{7}{2}$-$\frac{3}{2}$,$\frac{5}{2}$  &    291.200& 84.42&         347 $\pm$ 13\\
 & &      290.500& 84.60&         307 $\pm$ 20\\
  &  $\frac{3}{2}$,$\frac{5}{2}$-$\frac{3}{2}$,$\frac{3}{2}$ &    120.700&119.23&         141 $\pm$ 8\\
 & &      120.500&119.44&          190 $\pm$ 30\\
   &  $\frac{1}{2}$,$\frac{3}{2}$-$\frac{1}{2}$,$\frac{1}{2}$ &   270.200&163.12&         58 $\pm$ 9\\
 & &      269.800&163.40&          67 $\pm$ 10\\
\\
p-H$_2$O &4$_{31}$--3$_{22}$&      552.300& 56.33&          299 $\pm$ 26\\
 &4$_{22}$--3$_{13}$&              454.300& 57.64&          309 $\pm$ 42\\
 &7$_{26}$--6$_{15}$&              1021.00& 59.99&          136 $\pm$ 20\\
 &4$_{31}$--4$_{04}$&              552.300& 61.81&          314 $\pm$ 42\\
 &8$_{08}$--7$_{17}$&              1070.60& 63.46&          98 $\pm$ 31\\
 &3$_{31}$--2$_{20}$&              410.400& 67.09&          223 $\pm$ 18\\
 &5$_{24}$--4$_{13}$&              598.800& 71.07&          227 $\pm$ 17\\
 &7$_{17}$--6$_{06}$&              843.800& 71.54&          185 $\pm$ 14\\
 &6$_{06}$--5$_{15}$&              642.700& 83.28&          248 $\pm$ 26\\
 &3$_{22}$--2$_{11}$&              296.800& 89.99&          341$\pm$ 20\\
 &5$_{42}$--5$_{33}$&              877.800& 94.21&          67 $\pm$ 17\\
 &4$_{04}$--3$_{13}$&              319.500&125.35&          147 $\pm$ 6\\
 &3$_{31}$--3$_{22}$&              410.400&126.71&          32$\pm$ 4\\
 &3$_{13}$--2$_{02}$&              204.700&138.53&          188$\pm$ 6\\
 &4$_{13}$--3$_{22}$&              396.400&144.52&          61 $\pm$ 6\\
 &4$_{31}$--4$_{22}$&              552.300&146.92&          38 $\pm$ 7\\ 
 &3$_{22}$--3$_{13}$&              296.800&156.19&          137 $\pm$ 7\\
\\
o-H$_2$O&4$_{32}$--3$_{21}$&       550.400& 58.70&          452 $\pm$ 27\\
 &8$_{18}$--7$_{07}$&              1070.70& 63.32&          249 $\pm$ 39\\
 &7$_{16}$--6$_{25}$&              1013.20& 66.09&          305 $\pm$ 29\\
 &3$_{30}$--2$_{21}$&              410.700& 66.44&          468$\pm$ 16\\
 &3$_{30}$--3$_{03}$&              410.700& 67.27&          244 $\pm$ 15\\
 &7$_{07}$--6$_{16}$&              843.500& 71.95&          319$\pm$ 13\\
 &7$_{25}$--6$_{34}$&              1125.70& 74.95&          194 $\pm$ 28\\
 &3$_{21}$--2$_{12}$&              305.300& 75.38&          589 $\pm$ 24\\
 &4$_{23}$--3$_{12}$&              432.200& 78.74&          518$\pm$ 34\\
 &6$_{16}$--5$_{05}$&              643.500& 82.03&          460 $\pm$ 26\\
 &6$_{43}$--6$_{34}$&              1088.80& 92.81&          81 $\pm$ 11\\
 &4$_{41}$--4$_{32}$&              702.300& 94.71&          115$\pm$ 13\\ 
 &2$_{21}$--1$_{10}$&              194.100&108.07&          290$\pm$ 16\\
 &4$_{14}$--3$_{03}$&              323.500&113.54&          661 $\pm$ 24\\
 &4$_{32}$--4$_{23}$&              550.400&121.72&          48 $\pm$ 7\\
 &4$_{23}$--4$_{14}$&              432.200&132.41&          93 $\pm$ 5\\
 &5$_{14}$--5$_{05}$&              574.700&134.93&          54$\pm$ 8\\
 &3$_{30}$--3$_{21}$&              410.700&136.50&          100 $\pm$ 9\\
 &3$_{03}$--2$_{12}$&              196.800&174.63&          155 $\pm$ 6\\
 &2$_{12}$--1$_{01}$&              114.400&179.53&          179 $\pm$ 6\\
 &2$_{21}$--2$_{12}$&              194.100&180.49&          68 $\pm$ 6\\
\\
\textrm{[O I]}& $^3$P$_1$--$^3$P$_2$&  227.712&63.18& 6842$^{b}$ $\pm$ 165\\
\textrm{[O I]}&$^3$P$_0$--$^3$P$_1$&  326.579&145.53& 382$^{b}$ $\pm$ 21\\
\\
\textrm{[C II]} &$^2$P$_{3/2}$--$^2$P$_{1/2}$&   91.210& 157.74&   586$^{c}$ $\pm$ 15\\
\enddata
\tablecomments{The CO \& OH (84 $\mu$m) and CO \& o-H$_{2}$O (113 $\mu$m) emission lines are excluded due to line blending. All lines within the wavelengths from 95 $\mu$m to 102 $\mu$m are not included and are unreliable because of high noise levels.}
\tablenotetext{a}{{ }All line fluxes are extracted from the central spaxel, then they are corrected by the PACS/PSF.}
\tablenotetext{b}{{ }Flux measured from 15 spaxels of one nod position.}
\tablenotetext{c}{{ }Flux measured from 25 spaxels of one nod position.}
\end{deluxetable}

\clearpage
\begin{deluxetable}{cccc}
\tablewidth{0pt}
\tabletypesize{\scriptsize}
\scriptsize
\tablecaption{Summary of the Rotational Diagram Results \label{tbl_rot}}
\tablehead{\colhead{ Species }     &\colhead{Component}      &\colhead{\Trot} 
            &\colhead{\TotalN(molecule)$^{a}$}  \\
\colhead{   }     &\colhead{ }      &\colhead{(K)}      &\colhead{(10$^{46}$)}   }
\startdata
CO$^{b}$  &  warm  &  359 $\pm$ 4  &  373  $\pm$ 12\\ 
          &  hot   &  783 $\pm$ 32  &  77 $\pm$ 12\\ 
H$_{2}$O  & Para   & 171 $\pm$ 5 & 0.17 $\pm$ 0.006\\
          & Ortho  & 197 $\pm$ 2 & 0.24 $\pm$ 0.006\\
OH & $^2\Pi_{3/2}$--$^2\Pi_{3/2}$ & 193  $\pm$ 7 & 0.13 $\pm$ 0.008\\
   & $^2\Pi_{1/2}$--$^2\Pi_{1/2}$ & 128  $\pm$ 5  & 0.60 $\pm$ 0.10\\  
\enddata
\tablenotetext{a}{{ }The total number of molecules.}
\tablenotetext{b}{{ }CO ladders are fitted by two rotational temperatures with a break-point at \Eup{} = 1800 K.}
\end{deluxetable}

\clearpage
\begin{deluxetable}{cccccccc}
\tablecolumns{8}
\tablewidth{0pt}
\tabletypesize{\scriptsize}
\scriptsize
\tablecaption{Summary of the best-fit LVG model parameters \label{tbl_radex}}
\tablehead{
\colhead{} &
\multicolumn{3}{c}{One Component}  & \colhead{}&
\multicolumn{3}{c}{Power Law} \\
\cline{2-4} \cline{6-7} 
\colhead{ Species } &
\colhead{$T\rm_{kin}$} &
\colhead{$n(\rm H_2$)} &
\colhead{$N$(molecule)/$\Delta\textit{v}$} & 
\colhead{$\chi^2$} &
\colhead{power-law index, b}  &
\colhead{$n(\rm H_2$)} &
\colhead{$\chi^2$} \\
\colhead{} & \colhead{(K)} & \colhead{(cm$^{-3}$)} & \colhead{(cm$^{-2}$ km$^{-1}$s)} & \colhead{} &  & \colhead{(cm$^{-3}$)} &  
}
\startdata
CO & 5000 & 3.2 $\times$ 10$^{4}$ & 10$^{15}$ & 18.3 & 2.6 &  1.8 $\times$ 10$^{6}$ &  8.4 \\
 \hline
H$_{2}$O &  500    &  10$^{7}$  & 3.2 $\times$ 10$^{16}$ & 40.4 & 2.2 & 10$^{13}$ & 189.1   \\ 
 \hline 
OH & 175$^{a}$ &  1.3 $\times$ 10$^{9}$ & 1.8 $\times$ 10$^{15}$    &  8.1 \\
   & 125$^{b}$ &  1.0 $\times$ 10$^{8}$ & 1.0 $\times$ 10$^{15}$    &  15.5 \\
\enddata 
\tablenotetext{a}{{ }The best-fit model without IR-pumping effect.}
\tablenotetext{b}{{ }The best-fit model with IR-pumping effect.}
\end{deluxetable}

\clearpage
\begin{deluxetable}{cccccccc}
\tablewidth{0pt}
\tablecaption{Line luminosities in FIR$^{a}$ 
 \label{tbl_lum}}
\tablehead{
 \colhead{$L_\textrm{Continuum}$$^{b}$}
 &\colhead{$L_\textrm{FIR}$$^{c}$} 
 &\colhead{$L_\textrm{mol}$$^{d}$} 
 &\colhead{$L_\textrm{[O I]}$$$} 
 &\colhead{$L_\textrm{CO}$} 
 &\colhead{$L_\textrm{H$_{2}$O}$} 
 &\colhead{$L_\textrm{OH}$}      
 } 
\startdata
  530 &  1.00 &  0.68  &  0.32 &  0.21 &  0.36 &  0.11  \
\enddata
\tablenotetext{a}{{ }In units of 10$^{-2}$ \Lsun.}
\tablenotetext{b}{{ }Total FIR luminosity from the whole PACS spaxels.}
\tablenotetext{c}{{ }Total FIR line luminosity, $\textit{L}_\textrm{FIR}$ = $\textit{L}_\textrm{CO}$ + $\textit{L}_\textrm{H$_{2}$O}$ + $\textit{L}_\textrm{OH}$ + $\textit{L}_\textrm{[O I]}$.}
\tablenotetext{d}{{ }Molecular line luminosity, $\textit{L}_\textrm{mol}$ = $\textit{L}_\textrm{CO}$ + $\textit{L}_\textrm{H$_{2}$O}$ + $\textit{L}_\textrm{OH}$.}
\end{deluxetable}

\clearpage
\begin{deluxetable}{clcc}
\tabletypesize{\scriptsize}
\scriptsize
\tablewidth{0pt}
\tablecaption{Model parameters for the disk and envelope model \label{tbl_robitaille}}
\tablehead{
\colhead {Parameter} &\colhead{Description}  &\colhead{Unit}  
         &\colhead{Value}} 
\startdata
$M_{\ast}$ & Stellar mass   & \Msun     & 0.56\\
$R_{\ast}$ & Stellar radius & \Rsun     & 8.05 \\
$M_\textrm{env}$ & Envelope mass & \Msun      & 0.23 \\
$\dot{\textit{M}}_\textrm{env}$ & Envelope accretion rate & 10$^{-6}$ \Msun{} $yr^{-1}$ & 3.00 $^{a}$ \\
$R_\textrm{e,max}$ & Envelope outer radius & AU & 8472 \\
$M_\textrm{disk}$ & Disk mass & 10$^{-4}$ \Msun & 9.12  \\
$R_\textrm{d,max}$ & Disk outer radius (Centrifugal radius) & AU & 17.40 \\
$h_\textrm{100}$ & Disk scale height at 100 AU & AU & 6.11  \\
$\alpha$ & Disk radial density exponent &  & 2.08   \\
$\beta$ & Disk scale height exponent &  & 1.08   \\
$\theta_\textrm{cav}$ & Cavity opening angle & $^\circ$ & 5.21   \\
$A_\textrm{v}$ & Extinction &  & 12.19   \\
$\textit{L}_\textrm{tot}$ & Total luminosity & \Lsun & 10.00 $^{a}$   \\
$\theta_\textrm{incl}$ & Inclination & $^\circ$ & 41.41  \\ 
\enddata
\tablecomments{All parameters, except for the envelope accretion rate and the total luminosity, are obtained from Robitaille's online fitting tool (model ID : 3005892).}
\tablenotetext{a}{{ }The best-fit model parameters derived from our explored modeling using a fixed distance of 120 pc.}
\end{deluxetable}

\clearpage

\clearpage

\clearpage

\clearpage

\clearpage

\clearpage

\clearpage

\clearpage

\clearpage

\clearpage

\clearpage

\clearpage

\clearpage

\clearpage

\clearpage

\clearpage

\clearpage

\clearpage

\clearpage

\clearpage

\clearpage

\clearpage

\clearpage

\clearpage

\clearpage

\end{document}